\documentclass[12pt]{iopart}
\usepackage{graphicx}
\usepackage{caption}

\usepackage{multirow}
\usepackage{varioref}

\begin{document}

% Short title and long title
\title[UV LED Space Qualification]{Characterising and Testing Deep UV LEDs for Use in Space Applications}

\author{D Hollington, J T Baird, T J Sumner and P J Wass}

\address{High Energy Physics Group, Physics Department, Imperial College London, Prince Consort Road, London, SW7 2BW, UK.}
\ead{d.hollington07@imperial.ac.uk}

\begin{abstract}
Deep ultraviolet (DUV) light sources are used to neutralise isolated test masses in highly sensitive space-based gravitational experiments. An example is the LISA Pathfinder charge management system, which uses low-pressure mercury lamps. A future gravitational-wave observatory such as eLISA will use UV light-emitting diodes (UV LEDs), which offer numerous advantages over traditional discharge lamps. Such devices have limited space heritage but are are now available from a number of commercial suppliers. Here we report on a test campaign that was carried out to quantify the general properties of three types of commercially available UV LEDs and demonstrate their suitability for use in space. Testing included general electrical and UV output measurements, spectral stability, pulsed performance, temperature dependence as well as thermal vacuum, radiation and vibration survivability.

\end{abstract}

% See guide as to what is required here:
%\pacs{}
%\keywords{Min three, max seven keywords}
%\submitto{\cqg}

%%%%%%%%%%%%%%%%%%%%%%%%%%%%%%%%%%%%%%%%%%%%%%%%%%

\section{Introduction}

Due to launch in late 2015, LISA Pathfinder \cite{Armano2015} is a European Space Agency (ESA) precursor mission for a space-based gravitational wave observatory \cite{Amaro2012}. At the heart of the experiment are two cubic gold-platinum test masses, free-falling under gravity and each enclosed within their own capacitive inertial sensor. The gold-coated test masses act as mirrors for a laser interferometer used to measure changes in their separation with pico-metre accuracy. The spacecraft follows the motion of the test masses along the axis of their separation using micro-Newton thrusters while electrostatic forces are applied to control the motion in other degrees of freedom. To demonstrate the feasibility of detecting gravitational waves, the test masses need to maintain pure geodesic motion achieved by the elimination of all non-gravitational forces to a level below $6\times10^{-14}\,\textup{NHz}^{-1/2}$ in a frequency band between 1 and 10\,mHz.

Although the test masses are enclosed within the satellite there will be an inevitable build up of electric charge on the test masses due to incident ionising radiation from space \cite{Jafry1996}. A charged test mass will experience electrostatic forces through interaction with electric fields within the inertial sensor \cite{Weber2012}, as well as Lorentz forces via coupling with interplanetary magnetic fields \cite{Sumner2000}. Left unchecked, these forces can limit the performance of the instrument. They also introduce operational constraints and artefacts in the data \cite{Shaul2005} which make it necessary to control the  charge. While missions such as CHAMP \cite{Reigber2002}, GRACE \cite{Tapley2004},  GOCE \cite{Rummel2011} and MICROSCOPE \cite{Touboul2001} rely on a wire connection to ground to avoid charge build-up, the force noise goals of LISA Pathfinder necessitate a non-contact method to avoid disturbing the test masses.  The use of photoelectric emission from UV photons is such a method and was first demonstrated on the Gravity Probe B mission \cite{Buchman1995}.

LISA Pathfinder includes a charge management subsystem (CMS) \cite{Sumner2009} exploiting the photoelectric effect by using $253.7\,\textup{nm}$ light from low-pressure mercury discharge lamps to transfer charge between the gold-coated surfaces of the test mass and the surrounding housing and electrodes of the sensor. In its most simplistic form this means illuminating the housing and electrodes to add negative charge to the test mass or illuminating the test mass itself in order to remove negative charge. 

The capacitive inertial sensors of a future ESA gravitational wave observatory will be based on a similar design to those aboard LISA Pathfinder and as such will also require a charge management system \cite{Vitale2014} as will any other mission using isolated proof-masses, such as STEP \cite{Sumner2007}. Since the design and production of the LISA Pathfinder CMS, several types of deep ultraviolet light-emitting diodes (UV LEDs) have become commercially available with peak wavelengths below $260\,\textup{nm}$. These devices offer many advantages that make them promising candidates to replace mercury lamps as a CMS light source. However, unlike mercury lamps which have space heritage prior to the LISA Pathfinder mission \cite{Buchman1995, Adams1987}, such devices have seldom been used in space and never in an extended mission.

The aim of this work is to compare three types of commercially available UV LEDs and determine their suitability for use in space via a series of rigorous tests. A particular emphasis is placed on the unique and strict requirements needed to be fulfilled by any potential CMS light source.

%%%%%%%%%%%%%%%%%%%%%%%%%%%%%%%%%%%%%%%%%%%%%%%%%%

\section{LISA Pathfinder Test Mass Discharging}

Before considering what will be required of an improved light source for a generic CMS it is important to summarise the performance and functionality of the LISA Pathfinder CMS. The LISA Pathfinder CMS consists of three hardware parts, the main one being the UV Light Unit (ULU). It contains six programmable, customised Pen-Ray low-pressure mercury discharge lamps\footnote{http:\textbackslash\textbackslash uvp.com\textbackslash penraylightsources.html}, three for each inertial sensor, as well as the electronics required for their operation. Each lamp has an integrated heater to allow operation at low temperatures as well as an optics barrel to collect, filter and focus the emitted light into a fibre. A band-pass filter transmits the mercury line at $253.7\,\textup{nm}$ line responsible for photoemission while blocking light at $184.4\,\textup{nm}$, which is harmful to subsequent elements of the optical path and longer wavelength lines and continuum emission to avoid injecting unnecessary light into the sensor.

The other two hardware items are the Fibre Optic Harness (FOH) and Inertial Sensor UV Kit (ISUK). The FOH routes the light from the ULU in the outer compartment of the spacecraft towards each inertial sensor at the centre and consists of a series of custom made fibre optics with separate chains for each lamp. The UV light is finally injected into the sensors inside their vacuum enclosures via the ISUKs which are custom-made ultra-high vacuum fibre feed-throughs. There are three ISUKs for each sensor, two pointing at the housing and one at the test mass. Ideally, any future light source would be compatible with the LISA Pathfinder FOH and ISUK, allowing the same designs and materials to be reused.

The maximum UV power entering the sensor varies depending on the optical chain but is typically of order 100\,nW with a dynamic range of $\approx200$ achieved using a pulse-width modulation technique at kHz frequencies. Test mass discharging can be performed in two different modes: in fast discharge mode, the test mass is allowed charge up over several days until it reaches a level where charge-related disturbances become problematic, typically $\approx10^{7}$ elementary charges (e). The sensor is then illuminated at a relatively high UV power level to reduce the charge below $10^{5}$\,e over the course of 10 to 20 minutes. In continuous discharging mode a lower UV power level is used such that the discharging rate cancels the environmental charging rate, predicted to be $\approx10\,\textup{to}\,100\,\textup{es}^{-1}$ \cite{Araujo2005, Wass2005}, and the test-mass charge is maintained at a level below $10^5$\,e.

Each lamp requires its own high-voltage electronics to produce the $\approx600\,\textup{V}$ needed to initiate discharge in the mercury vapour, falling to around half this level during operation. The typical power consumption of a mercury lamp at full power is $\approx4\,\textup{W}$. When initially switched on, the rise-time of the UV output is temperature dependent where the time for the output to rise from 5\% to 95\%  of its maximum value ranges from $\approx50\,\textup{s}$ at $20\,^{\circ}\textup{C}$ to $\approx15\,\textup{s}$ at $40\,^{\circ}\textup{C}$. The vapour pressure of mercury varies with temperature and therefore UV power emitted by a lamp at a particular operational setting is also temperature-dependent. Related to the vapour pressure, at temperatures below $15\,^{\circ}\textup{C}$ it is necessary to pre-warm lamps prior to operation. While lamp output will degrade during the LISA Pathfinder mission lifetimes are predicted to be over $1000\,\textup{hours}$ of continuous use at high output power.

\subsection{Complications}

There are three complications to the simple concept of discharging presented so far. The first is the absorption of UV light by unintended surfaces. For example, although the light may be directed at one surface, say the test mass, light is inevitably reflected leading to absorption on the sensor housing or electrodes opposite, generating a photocurrent acting against the intended direction of discharge. At $253.7\,\textup{nm}$, the reflectivity of the gold coated surfaces is 36\% at normal incidence but increases rapidly for incident angles shallower than $45^{\circ}$ \cite{Johnson1972}. In the inertial sensor,  most light is incident at between 45 and 70$^{\circ}$ relative to the surface normal and as such, significant amounts of light will be absorbed by secondary surfaces.

A second complication arises due to unavoidable contamination of the gold surfaces inside the inertial sensor. The work function of pure gold, deposited in vacuum and measured \textit{in situ} is $5.2\,\textup{eV}$ \cite{Huber1966}. Upon exposure to air, adsorbates reduce the effective work function to as low as $4.2\,\textup{eV}$ \cite{Saville1995}, with water and hydro-carbons having particular influence. This surface contamination persists even when the sample is placed in vacuum and after modest baking \cite{Hechenblaikner2012}. Given that the energy of the usable photons from mercury lamps is $4.89\,\textup{eV}$, it is only through surface adsorbates that discharge can occur at all. However, relying on inherently uncertain surface properties leads to nominally identical sensor surfaces having significantly different photoelectric properties. Studies have found that the quantum yield, the number of emitted photoelectrons per absorbed photon, of gold can vary from $10^{-6}$ to $10^{-4}$ \cite{Schulte2009}. When combined with the distribution of absorbed light due to reflections, in the most extreme case a significant asymmetry in yield can prevent discharging in one direction no matter which surface is initially illuminated.

The final complication is the presence of local electric fields within the inertial sensor. Both alternating and direct current (AC and DC) voltages can be applied to the 18 separate electrodes surrounding the test mass to enable the test mass position sensing and actuation \cite{Weber2007}. The voltage used to bias the test mass for capacitive sensing is a $100\,\textup{kHz}$ sine wave with a nominal $5.4\,\textup{V}$ amplitude applied to six of the electrodes in the sensor. This so called injection bias results in a test mass potential with respect to the grounded electrode housing of $\pm0.6\,\textup{V}$. Sinusoidal actuation voltages are also applied to the remaining 12 position-sensing electrodes at a range of audio frequencies with amplitudes of up to $7\,\textup{V}$. The potential difference between test mass and sensor therefore varies greatly depending on location. Crucially, the energy of the emitted photoelectrons is less than $1\,\textup{eV}$ meaning that photocurrents are strongly influenced by these electric fields, to the point where they can be completely suppressed. Although avoided as it introduces electrostatic forces in the measurement bandwidth, DC voltages may also be applied to individual electrodes to aid discharging.

%%%%%%%%%%%%%%%%%%%%%%%%%%%%%%%%%%%%%%%%%%%%%%%%%%

\section{UV LEDs}

Using UV LEDs as a CMS light source has several obvious advantages over the mercury lamps used for LISA Pathfinder. They offer significant mass and volume savings not only because of their smaller size but also because of a reduction in the complexity of the associated electronics and optics. Electrical power consumption is significantly reduced and, depending on how they are operated, they offer a higher range of UV output power and longer lifetimes than mercury lamps.

UV LEDs are available that can produce light with a wavelength $<254\,\textup{nm}$ and therefore photons of higher energy than the mercury spectral line. Moving away from the work function of contaminated gold not only should the quantum yield increase but at a photon energy $>5.2\,\textup{eV}$ it is possible to liberate electrons from pure gold. This opens the possibility of reducing or removing surface contamination, for example by aggressive baking under vacuum or ion sputtering, without the fear of preventing discharging. Without the unpredictable photoelectric properties of a contaminated surface, the risk of bipolar discharging not being possible would be removed. A parallel study has involved measuring the quantum yield from a number of prepared surfaces and this will be reported separately \cite{Wass2015}.

Being semiconductor devices, it is possible to pulse UV LEDs at high frequencies and synchronise them with the AC voltages present in the inertial sensor \cite{Sun2006}. Switching the device on only when the electric fields in the region under illumination enhance the desired direction of discharge elegantly mitigates the problem of asymmetric photoelectric properties as the photo-current would be able to flow in one direction while being partially or completely suppressed in the opposite, undesirable direction. This method also provides redundancy as shifting the phase of the pulses by $180^{\circ}$ would reverse the direction of net photocurrent and allow a single point of UV injection to discharge both positively and negatively. Furthermore, the discharge rate could be tuned by adjusting the phase of the UV pulses with respect to the synchronising voltage, increasing the dynamic range of the discharge system. Finally, synchronous discharging would be more efficient compared to a DC scheme where for half of the injection bias cycle the voltages act against the desired direction.

With the potential advantages clear, work has already been carried out by other groups to demonstrate that UV LEDs are suitable for space applications. These have mainly focused on one particular type of device with a peak wavelength at approximately $255\,\textup{nm}$ supplied by Sensor Electronic Technology (SET) \cite{SETi2015}. These tests have yielded promising results suggesting that this type of device has impressive lifetimes, can be pulsed at high frequencies and are radiation hard \cite{Sun2006, Sun2009}. However, pulsed discharging has only been demonstrated at frequencies of $1\,\textup{kHz}$ and $10\,\textup{kHz}$ with $50\%$ duty cycles. The obvious electric field with which to to synchronise in the inertial sensor would be the $100\,\textup{kHz}$ injection voltage, with duty cycles $<50\%$ resulting in pulses of $<5\,\mu\textup{s}$ duration. In addition, previous work has not focussed on dynamic range or presented a systematic study on a number of devices.

More recently, shorter wavelength devices from the same supplier have come to market which nominally peak at $240\,\textup{nm}$. This shorter wavelength could offer additional potential benefits for discharging. More recently still, devices which peak at $250\,\textup{nm}$ and supplied by Crystal IS (CIS) have appeared \cite{CIS2015}. For space applications a second source supplier can be crucial.

Both types of device (SET and CIS) consist of a structure of AlGaN layers grown on a substrate via patented chemical vapour deposition processes. Broadly speaking the ratio of aluminium and gallium within the $Al_{x}Ga_{1-x}N$ alloy determines the peak wavelength of the device with increased aluminium content leading to lower wavelengths, \cite{Hirayama2005}. For the SET devices sapphire substrates are used while CIS use aluminium nitride, which they claim produces fewer lattice defects, leading to superior lifetimes and higher output powers. Both suppliers offer devices surface mounted or pre-packaged with integrated optics and according to their data sheets offer UV output powers of $\approx100\,\mu\textup{W}$ at electrical powers of $<100\,\textup{mW}$. Both also have wide operating temperature ranges of $-30\,^{\circ}\textup{C}$ to $+50\,^{\circ}\textup{C}$.

Despite the promise of UV LEDs as a future CMS light source several unanswered questions remain. Unlike mercury lamps, which have a space heritage pre-dating the LISA Pathfinder mission, UV LEDs have very limited short-term space exposure. Before a future CMS can be designed as a whole a precise understanding of the properties of its light source is required. Survivability needs to be demonstrated under environmental stresses such as thermal vacuum cycling, radiation exposure, vibration and shock.  Furthermore, a detailed understanding of the spectral properties of a device under a variety of operational scenarios, including any changes with operational temperature, drive current, pulse width, age or after irradiation is vital as it directly determines the yield and energy distribution of the photoelectrons emitted from a surface. Both the electrical properties and UV output of the devices under consideration are known to be strongly temperature dependent and this needs to be fully understood. 

Tests needs to be carried out against a clear requirement specification; the outcome of a recent focussed ESA technology study has been used here to define the range and extent for this work \cite{Sumner2015}. Testing was carried out on a range of commercially available UV LEDs that aimed to quantify and compare a variety of device characteristics including their general electrical and UV output properties, spectral stability, pulsed performance, temperature dependence as well as thermal vacuum, radiation and vibration survivability. It was equally important to test the underlying technologies such that if and when even shorter wavelength devices are released in future critical issues can be identified.

%%%%%%%%%%%%%%%%%%%%%%%%%%%%%%%%%%%%%%%%%%%%%%%%%%

\section{Test Devices}

In total, nine devices of three different types were tested. Two types of device were supplied by Sensor Electronic Technology (SET) with part references UVTOP-255-TO39-BL and UVTOP-240-TO39-HS while the other was supplied by Crystal IS (CIS) with a part reference OPTAN250J. The SET devices of the same type were produced from the same wafer, while the CIS devices were selected from a bin of devices with similar output properties as tested by the manufacturer. All the devices were packaged in TO-39 cans and identified by their nominal peak wavelength. The two SET device types had nominal peak wavelengths of $255\,\textup{nm}$ and $240\,\textup{nm}$ while the CIS device was $250\,\textup{nm}$. The SET-255 devices were sold as `research grade' and had an integrated ball lens focusing their output to a spot approximately $2\,\textup{mm}$ in diameter at a distance of about $20\,\textup{mm}$. The SET-240 devices had a hemispherical lens creating a beam with a typical $6^\circ$ spread, while the CIS-250 devices also had an integrated ball lens with a slightly shorter focal length than that of the SET device. According to the manufacturer the CIS-250 devices had received a $48\,\textup{hour}$ “burn-in” at $100\,\textup{mA}$ prior to delivery which was not the case for the SET devices. All the devices remained completely untouched and in secure storage prior to testing.

\begin{figure}[h]
\centering
\begin{minipage}[c]{0.35\textwidth}
\includegraphics[width=1.0\textwidth]{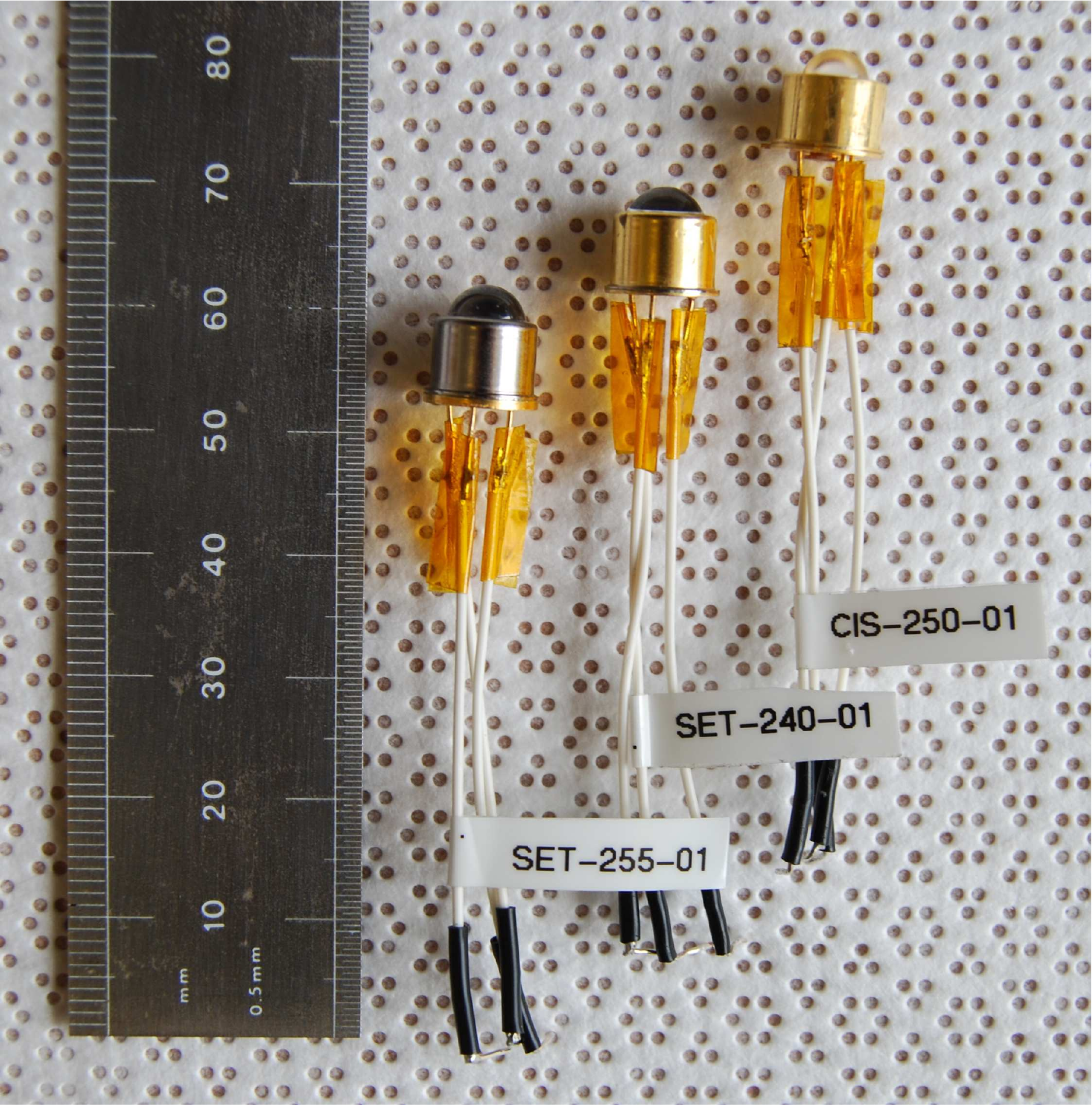}
\end{minipage}
\hfil
\begin{minipage}[c]{0.40\textwidth}
\caption[Photograph of devices with flying leads attached.]{\label{fig:DevicePhotos} From left to right the devices are SET-255-01, SET-240-01 and CIS-250-01. Each device has soldered flying leads attached, with the joint insulated with Kapton tape. Note that each device has a different type of lens while the scale on the left is in mm.}
\end{minipage}
\end{figure}

Before each device was tested it underwent an initial inspection where it was weighed, photographed and checked visually. It then had flying leads attached and was assigned a unique label and stored in an individual Electrostatic Discharge (ESD) bag. Three of the nine devices are shown in Figure \ref{fig:DevicePhotos}.

\subsection{Testing}

Testing was carried out from June to October 2014 and other than radiation and vibration, which were performed at external facilities, all tests were carried out with custom-made test equipment developed and built at Imperial College. All the work described here was done under ESA contract and included a high level of documentation and traceability with formal test procedures and task sheets. Data processing and analysis were carried out with the LISA Technology Package Data Analysis (LTPDA) toolbox \cite{LTPDA2015}.

Unless otherwise stated, the tests were performed with the device under test (DUT) fixed in a copper mount within a light sealed enclosure. The mount temperature was controlled using a thermoelectric system with two thermistors embedded within the mount (Peltier: DA-014-12-02, Controller: PR-59, both supplied by Laird Technologies). During low temperature testing the enclosure was flooded with nitrogen to prevent condensation and/or ice forming. A calibrated Hamamatsu S1337-1010BQ UV photodiode with a $1\,\textup{cm}^2$ sensitive area could be fixed opposite the DUT, in the same mount, capturing the total UV output of the device. Alternatively, with the photodiode removed, the output from the DUT could be injected into a $1\,\textup{mm}$ diameter, UV transparent fibre. Within the enclosure, the fibre itself was mounted on a 3-axis translation stage to position the fibre tip for optimal UV acceptance. Depending on the DUT, it was possible to couple $10\%$ to $20\%$ of the total light emitted into the fibre. The fibre was routed out of the enclosure through a rubberised seal and interfaced with either a spectrometer or a Hamamatsu H6780-06 photomultiplier tube (PMT).

Spectral measurements were made using a customized Princeton Instruments Model VM-502 spectrometer with slit widths chosen to give a resolution of $\approx0.3\,\textup{nm}$. The spectrometer diffraction grating was positioned with a stepper motor and its calibration was checked on a weekly basis against the well-defined position of spectral lines from an unfiltered mercury lamp source. When used in conjunction with a broadband deuterium lamp the spectrometer could also be operated as a monochromator. This made it possible to use a calibrated power meter (Sensor: 918D-UV-OD3R, Meter: 841-P-USB, both supplied by Newport) to cross-calibrate the absolute response as a function of wavelength of the photodiodes and PMT used during testing with a systematic uncertainty of $\pm2\%$. The temperature dependence of both the response and dark current for the photodiodes was measured over a temperature range from $-10\,^{\circ}\textup{C}$ to $+60\,^{\circ}\textup{C}$. Over this range the response was found to vary by less than $0.1\%$, the limit of the measurement. The dark current was found to increase exponentially with temperature measuring $0.01\,\textup{nA}$ at $-10\,^{\circ}\textup{C}$ and $1\,\textup{nA}$ at $+60\,^{\circ}\textup{C}$. Although this was small compared to the typical $\mu\textup{A}$ photodiode signals measured it was nonetheless subtracted during runs where a dark reading could be made.

Great care was taken throughout testing to comply with handling and usage recommendations for each device given on their data sheet and with ESA European Cooperation for Space Standardisation (ECSS) guidelines \cite{ECSS2009}. Precautions for handling ESD-sensitive devices were taken and UV LEDs were only handled wearing clean nitrile gloves, while also taking care to avoid mechanical shocks. Appropriate heat sinking was also used when soldering devices. A $10\,\textup{mA}$ maximum average drive current was chosen for the devices, well within datasheet recommendations. At this level, UV output powers at the end of a fibre were predicted to be $>10\,\mu\textup{W}$, several times higher than that delivered by the LISA Pathfinder CMS and consistent with the requirement specification \cite{Sumner2015}. A Keithley 2602A source meter was used to supply the devices with DC current. To drive the devices in a pulsed mode, a Tektronix AFG 3101 signal generator was used to trigger a custom-made set of electronics producing a variable current-limited pulse amplitude. A Tektronix AM503 current probe and an Agilent Infiniium MS09254A oscilloscope were used to study the pulsed behaviour.

The mainly automated laboratory-based tests were carried out in a sequential order taking a week per device. Once all  devices had been characterised in the laboratory, radiation testing was performed followed by vibration and shock, again in accordance with \cite{Sumner2015}. At several points in the test campaign a reference test was performed on each device. This consisted of a small subset of performance tests aimed at giving a baseline against which any potential changes in characteristics could be monitored. Reference measurements was performed on each device before and after the laboratory based tests, after the radiation test and after the vibration test.

%%%%%%%%%%%%%%%%%%%%%%%%%%%%%%%%%%%%%%%%%%%%%%%%%%

\section{Results}

The laboratory based tests were carried out on each device individually and are collated here to allow comparison.

\subsection{Initial Reference Characteristics}

The reference measurement consisted of three parts; a current-voltage (IV) scan performed while simultaneously measuring the total UV output of the DUT, a spectral scan while the DUT was driven at $1\,\textup{mA}$ DC, and a waveform measurement while the DUT was pulsed at $100\,\textup{kHz}$, $10\%$ duty cycle with a $10\,\textup{mA}$ drive current amplitude. The IV scan was done between $0\,\textup{mA}$ and $10\,\textup{mA}$, in steps of $0.1\,\textup{mA}$, dwelling at each setting for $1\,\textup{second.}$ All measurements were made with the DUT sealed in the air filled temperature controlled enclosure, held at $20\,^{\circ}\textup{C}$.

\begin{figure}[h]
\begin{minipage}[t]{0.5\textwidth}
\centering
\includegraphics[width=1.0\textwidth]{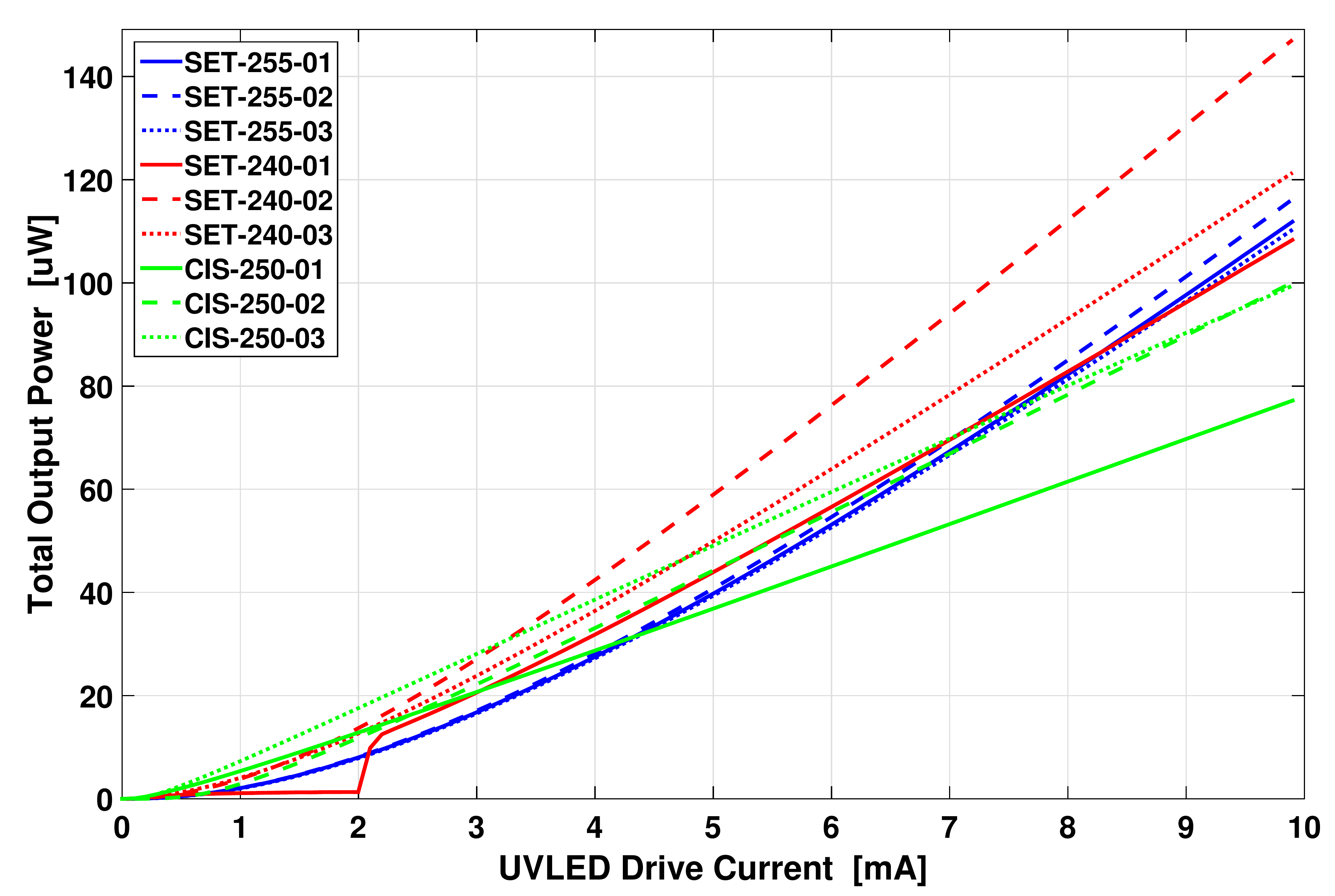}
\includegraphics[width=1.0\textwidth]{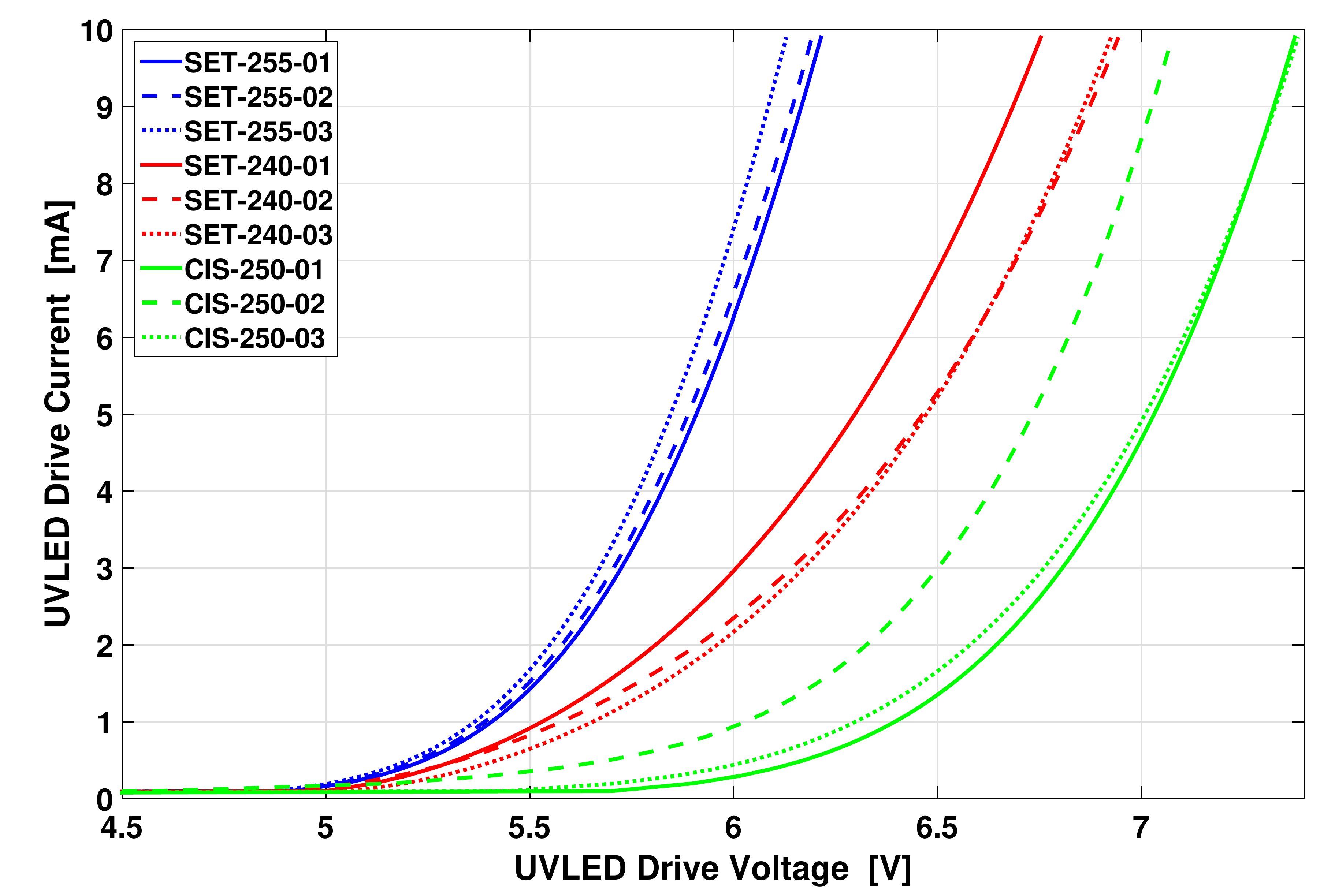}
\end{minipage}
\begin{minipage}[t]{0.5\textwidth}
\includegraphics[width=1.0\textwidth]{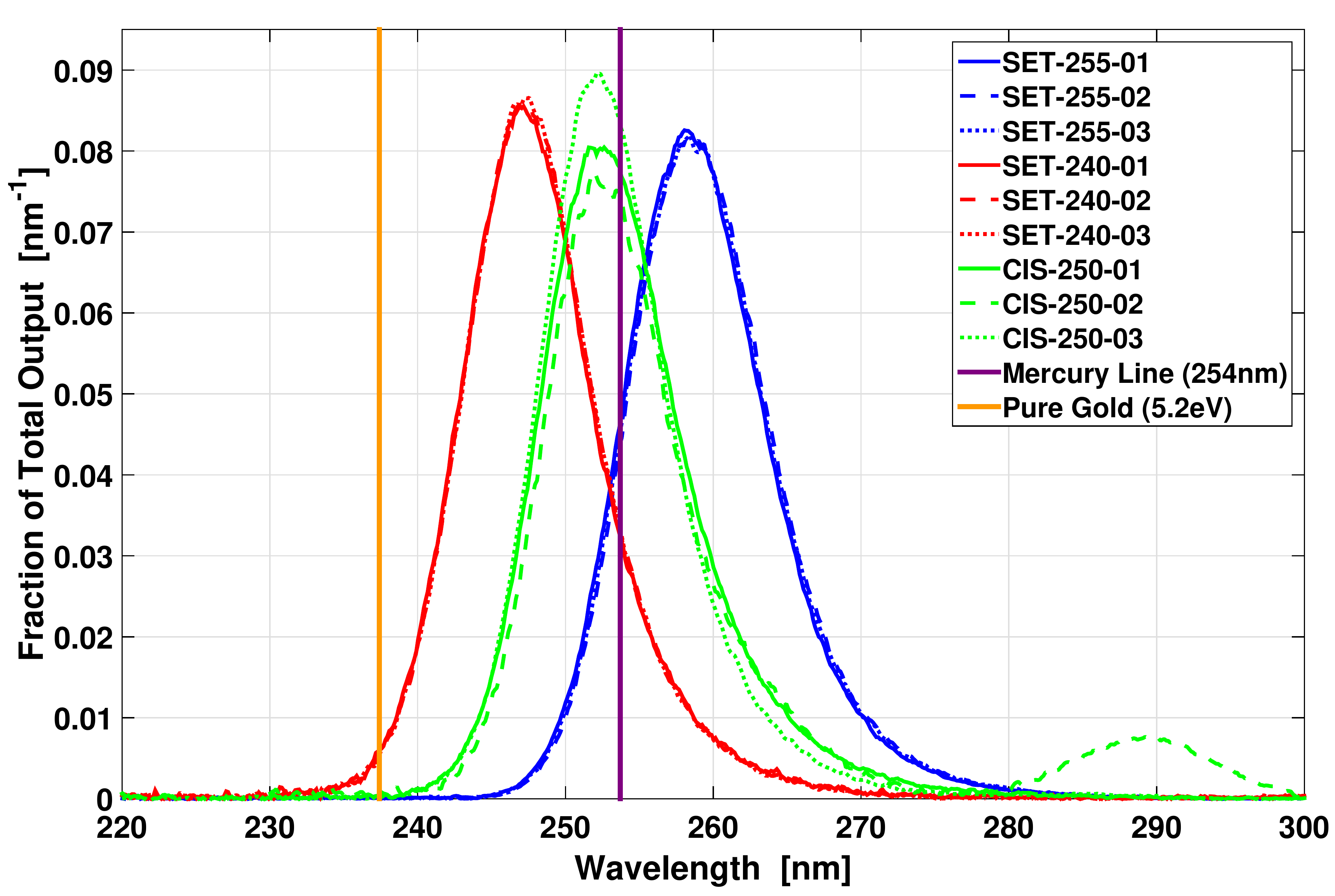}
\caption[Initial device characteristics.]{\label{fig:DeviceChar} Clockwise from bottom left: IV scans, simultaneously measured total UV output power and spectral scans. Note the unexpected jump at $2\,\textup{mA}$ during the SET-240-01 UV measurement. The cause of this anomaly is not clear but no such jump appears in the IV curve and the behaviour was not seen again in any other test.}
\end{minipage}
\end{figure}

Referring to Figure \ref{fig:DeviceChar}, all the initial IV scans show broad agreement with device data sheets and the three device types fall within a similar range, approximately $5\,\textup{V}$ to $7.5\,\textup{V}$ at $10\,\textup{mA}$. Note that the nine curves are approximately grouped by device type but there is a spread within each of these groups, the SET-255s showing the least variation. The simultaneously measured total UV output shows seven out of the nine devices had a maximum output at $10\,\textup{mA}$ of between $100\,\mu\textup{W}$ and $120\,\mu\textup{W}$. However, there are two devices that lie outside this range at $78\,\mu\textup{W}$ and $145\,\mu\textup{W}$. While keeping in mind the small sample size, the SET-240s produce the highest output while the CIS-250s have the lowest, at least at drive currents above $5\,\textup{mA}$. At the lowest current setting tested ($0.1\,\textup{mA}$) the devices emitted between $15\,\textup{nW}$ and $80\,\textup{nW}$. This equates to dynamic ranges of between $10^{3}$ and $10^{4}$, even in this simple operational scenario. It is also worth noting that the output from the CIS-250 devices appears slightly more linear with drive current than for the SET devices. The typical electrical power consumptions are in the tens of mW range leading to conversion efficiencies of $0.1\%$ to $0.2\%$. 

Figure \ref{fig:DeviceChar} also shows the measured spectrum of each device, normalised to have equal areas. The positions of the mercury $253.7\,\textup{nm}$ spectral line and the wavelength equivalent to the work function of pure gold are indicated for comparison. The spectra produced by SET devices of a particular type are almost indistinguishable from one another. While this is not unexpected as all devices were manufactured from the same wafer, it demonstrates the reproducibility of the measurement given that each scan was taken weeks apart. The CIS-250 devices show noticeable differences with CIS-250-02 standing out in particular with an unexpected secondary peak at $290\,\textup{nm}$ which accounts for approximately $10\%$ of the light emitted. 

A quantitative analysis was performed on each spectrum in order to extract both the peak wavelength and full width at half maximum (FWHM). The peak position was obtained by fitting a Gaussian function to the 50 points around the central maximum while ignoring the tails and the FWHM by finding the point at which the measured spectrum crossed (up/down) half its maximum value, interpolating between the two consecutive points where this occurs on either side. The results are summarised in Table \ref{tab:DeviceInitialSpecTable}.

\begin{table}[h]
\begin{center}
\begin{tabular}{ | c | c | c | c | c |}
\hline
\multirow{2}{*}{Device}   &  Peak Wavelength                                       &  FWHM                                       &  \multicolumn{2}{|c|}{Fraction Below}                       \\
                                          &  (nm)                                                            &  (nm)                                          &  $253.7\,\textup{nm} $  &  $237.4\,\textup{nm}$   \\
\hline
  SET-255-01                     &  $258.34 \pm 0.06$                                     &  $10.88 \pm 0.02$                     &  $0.13$                            &  $0.00$                            \\
  SET-255-02                     &  $258.63 \pm 0.06$                                     &  $11.01 \pm 0.01$                     &  $0.11$                            &  $0.00$                            \\
  SET-255-03                     &  $258.53 \pm 0.06$                                     &  $10.96 \pm 0.03$                     &  $0.12$                            &  $0.00$                            \\
  SET-240-01                     &  $247.20 \pm 0.06$                                     &  $10.11 \pm 0.04$                     &  $0.86$                            &  $0.01$                            \\
  SET-240-02                     &  $247.16 \pm 0.06$                                     &  $10.07 \pm 0.12$                     &  $0.85$                            &  $0.01$                            \\
  SET-240-03                     &  $247.15 \pm 0.06$                                     &  $10.21 \pm 0.16$                     &  $0.86$                            &  $0.01$                            \\
  CIS-250-01                     &  $252.52 \pm 0.06$                                     &  $10.71 \pm 0.07$                     &  $0.53$                            &  $0.00$                            \\
  CIS-250-02                     &  $252.48 \pm 0.06$                                     &  $10.31 \pm 0.16$                     &  $0.45$                            &  $0.00$                            \\
  CIS-250-03                     &  $252.20 \pm 0.06$                                     &  $9.58   \pm 0.11$                     &  $0.57$                            &  $0.00$                            \\
\hline
\end{tabular}
\caption[Device's initial spectral properties.]{\label{tab:DeviceInitialSpecTable} Initial spectral properties, measured at $20\,^{\circ}\textup{C}$ while driven at $1\,\textup{mA}$ DC. Also shown is the fraction of each spectrum below the mercury $253.7\,\textup{nm}$ spectral line and the wavelength equivalent of the pure gold work function ($237.4\,\textup{nm}$). Note the error  in the peak wavelength is dominated by the uncertainty in the spectrometer calibration.}
\end{center}
\end{table}

All three device types have peak wavelengths higher than their nominal values of $255\,\textup{nm}$, $240\,\textup{nm}$ and $250\,\textup{nm}$, though within the limits specified by the suppliers, up to $+7\,\textup{nm}$ for the SET devices and $+2.5\,\textup{nm}$ for CIS. While all three device types produce significant amounts of light at a wavelength shortwards of the $253.7\,\textup{nm}$ used in the LISA Pathfinder CMS, only the SET-240s produce light at an energy greater than the work function of pure gold, albeit only 1\% of their total.

\subsection{Spectral Stability}

Each device spectrum was remeasured under a variety of operational scenarios, including various temperatures ($-10\,^{\circ}\textup{C}$, $+20\,^{\circ}\textup{C}$ and $+40\,^{\circ}\textup{C}$, driven at $1\,\textup{mA}$ DC), DC drive currents ($1\,\textup{mA}$, $5\,\textup{mA}$ and $10\,\textup{mA}$, held at $20\,^{\circ}\textup{C}$) and pulsed duty cycle (5\%, 25\% or 50\%, at $100\,\textup{kHz}$, $20\,\textup{mA}$ amplitude). Including the reference measurements carried out before and after laboratory testing, ten measured spectra for each device were available for comparison and to check spectral stability under different operational conditions. Each spectrum was first quantified by extracting the spectral peak and FWHM as described previously. These data were then plotted for each device against the operational condition of interest, for example temperature, allowing the observation of any significant trends in the data. 

Within the uncertainty of the measurement, the spectra (peak position and FWHM) of all devices was found to be stable with DC drive current and pulsed duty cycle, the one exception being CIS-250-02. It was found that with increasing drive current, whether DC or pulsed, the amplitude of the secondary peak diminished relative to the primary. At $10\,\textup{mA}$ it made up $<1\%$ of the total emitted light compared to $10\%$ at $1\,\textup{mA}$. However, the main peak showed no measurable variation in terms of peak position or FWHM. 

\begin{figure}[h]
\centering
\includegraphics[height=0.25\textheight]{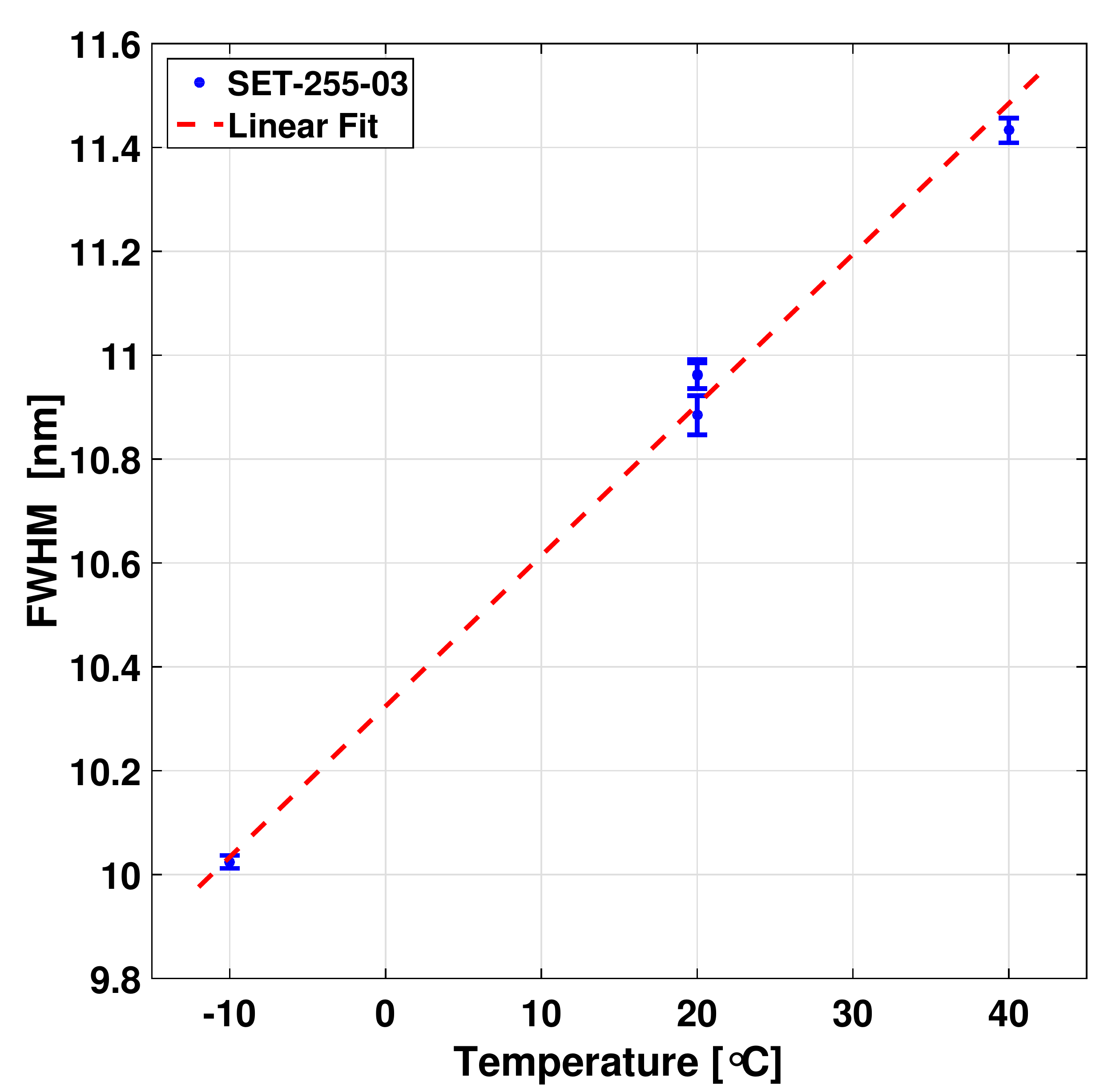}
\hfil
\includegraphics[height=0.25\textheight]{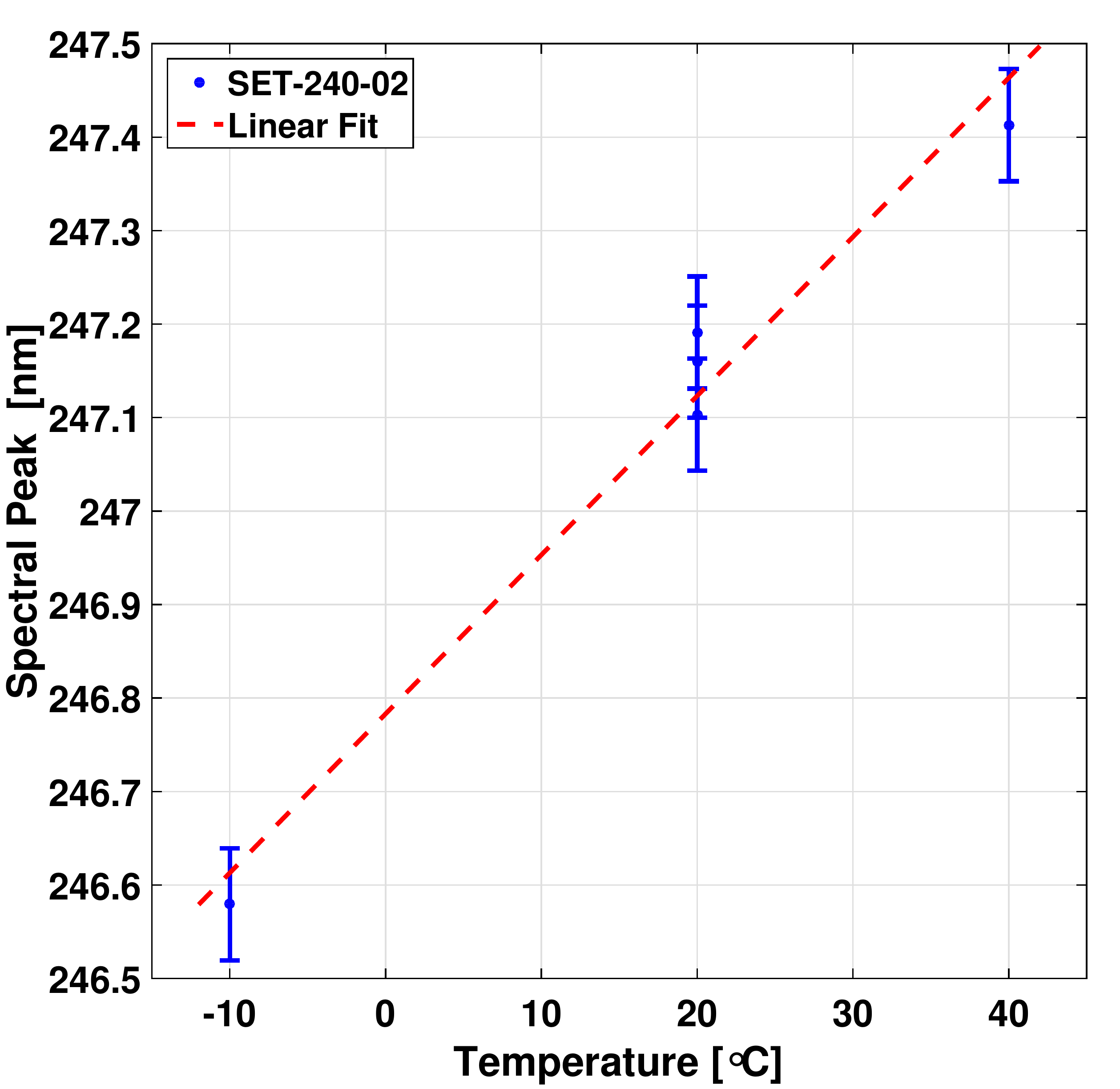}
\caption[Linear spectral properties.]{\label{fig:LinSpecProp} Left: Change in FWHM with temperature for SET-255-03. Right: Change in spectral peak position with temperature for SET-240-02.}
\end{figure}

Temperature was found to affect the FWHM of all devices as well as the peak position of the lower wavelength devices. In both cases the relationship was linear over the temperature ranged studied and two typical examples are shown in Figure \ref{fig:LinSpecProp}.

Linear fits were performed on each set of FWHM and spectral peak data with the results summarized in Table \ref{tab:DeviceLinFitSpec}. The three device types produced consistent results within each group. For the SET-255 and SET-240 devices a $1.5\,\textup{nm}$ reduction in FWHM was observed over the $50\,^{\circ}\textup{C}$ temperature range studied, while for the CIS-250 devices it was $1.0\,\textup{nm}$. While no effect was observed for the SET-255 devices within the uncertainty of the measurement ($\approx0.002\,\textup{nm}/^{\circ}\textup{C}$), the other two types also saw a shift in their spectral peak position of approximately $0.75\,\textup{nm}$ over the same temperature range.

\begin{table}[h]
\begin{center}
\begin{tabular}{ | c | c | c | c | c | }
\hline
\multirow{3}{*}{Device}  &  \multicolumn{2}{|c|}{FWHM Linear Fit}                                                &  \multicolumn{2}{|c|}{Spectral Peak Linear Fit}                                            \\
                                         &  Gradient                                                             &  Intercept                &  Gradient                                                             &  Intercept                        \\
                                         &  ($\textup{nm}/^{\circ}\textup{C}$)                &  (nm)                        &  ($\textup{nm}/^{\circ}\textup{C}$)                &  (nm)                                \\
\hline
  SET-255-01                     &  $0.031 \pm 0.003$                                            &  $10.30 \pm 0.07$  &  $-$                                                                      &  $-$                                  \\
  SET-255-02                     &  $0.024 \pm 0.004$                                            &  $10.5   \pm 0.1  $  &  $-$                                                                      &  $-$                                  \\
  SET-255-03                     &  $0.028 \pm 0.002$                                            &  $10.34 \pm 0.04$  &  $-$                                                                      &  $-$                                  \\
  SET-240-01                     &  $0.030 \pm 0.001$                                            &  $9.50   \pm 0.02$  &  $0.016 \pm 0.002$                                             &  $246.87 \pm 0.05$        \\
  SET-240-02                     &  $0.030 \pm 0.002$                                            &  $9.53   \pm 0.04$  &  $0.017 \pm 0.002$                                             &  $246.78 \pm 0.04$        \\
  SET-240-03                     &  $0.029 \pm 0.005$                                            &  $9.7     \pm 0.1  $  &  $0.017 \pm 0.002$                                             &  $246.80 \pm 0.04$        \\
  CIS-250-01                     &  $0.017 \pm 0.004$                                            &  $10.2   \pm 0.1  $  &  $0.009 \pm 0.002$                                             &  $252.29 \pm 0.06$        \\
  CIS-250-02                     &  $0.015 \pm 0.004$                                            &  $9.9     \pm 0.1  $  &  $0.011 \pm 0.002$                                             &  $252.2   \pm 0.1  $        \\
  CIS-250-03                     &  $0.022 \pm 0.002$                                            &  $9.3     \pm 0.1  $  &  $0.013 \pm 0.001$                                             &  $251.90 \pm 0.03$        \\
\hline
\end{tabular}
\caption[Linear fits of spectral properties with temperature.]{\label{tab:DeviceLinFitSpec} A summary of the linear fits found for each device describing the change in their spectral properties with temperature. Note that no significant temperature dependence was found for the position of the SET-255 spectral peaks.}
\end{center}
\end{table}

\subsection{Pulsed Capabilities}

\begin{figure}[h]
\centering
\includegraphics[width=0.49\textwidth]{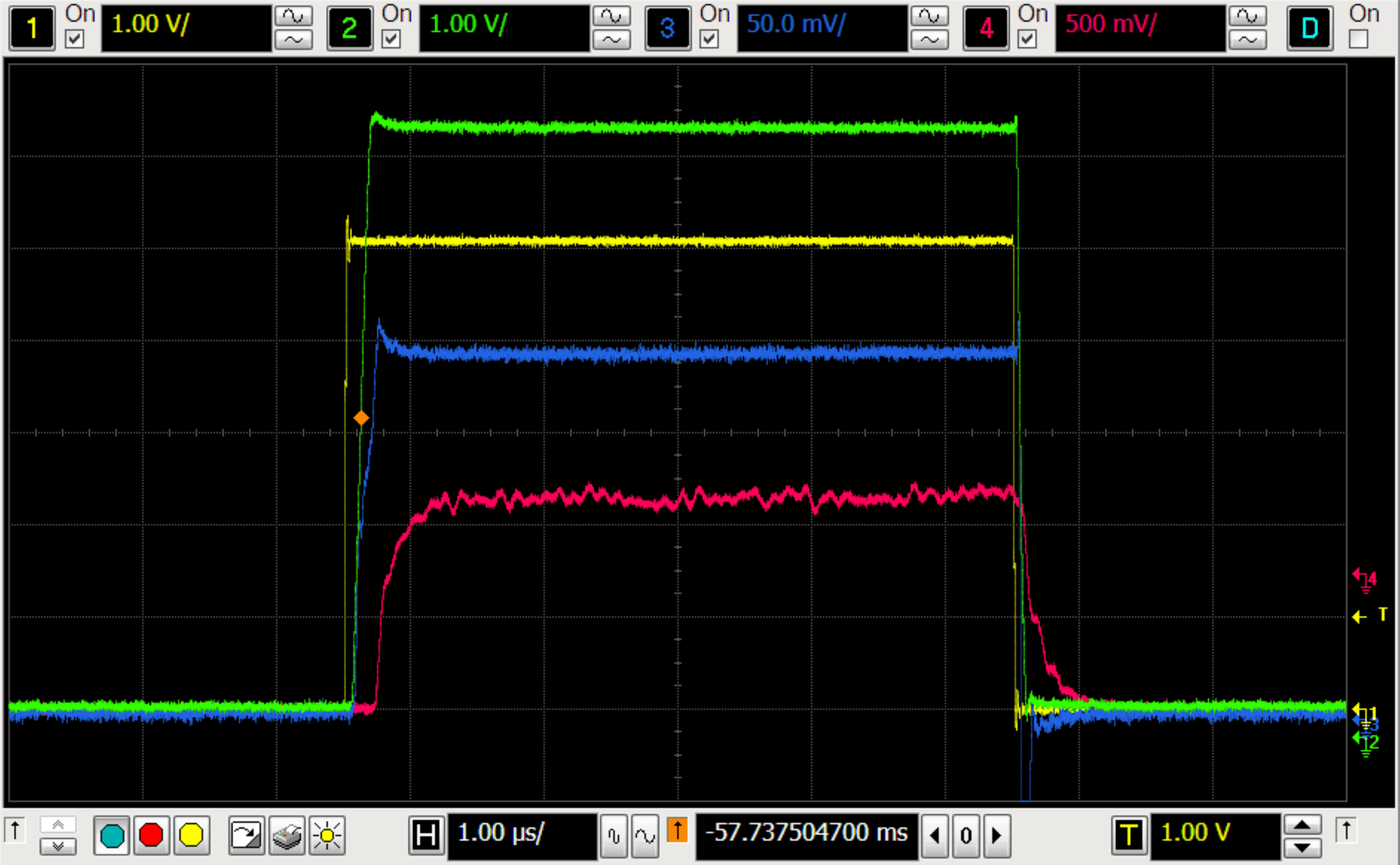}
\includegraphics[width=0.49\textwidth]{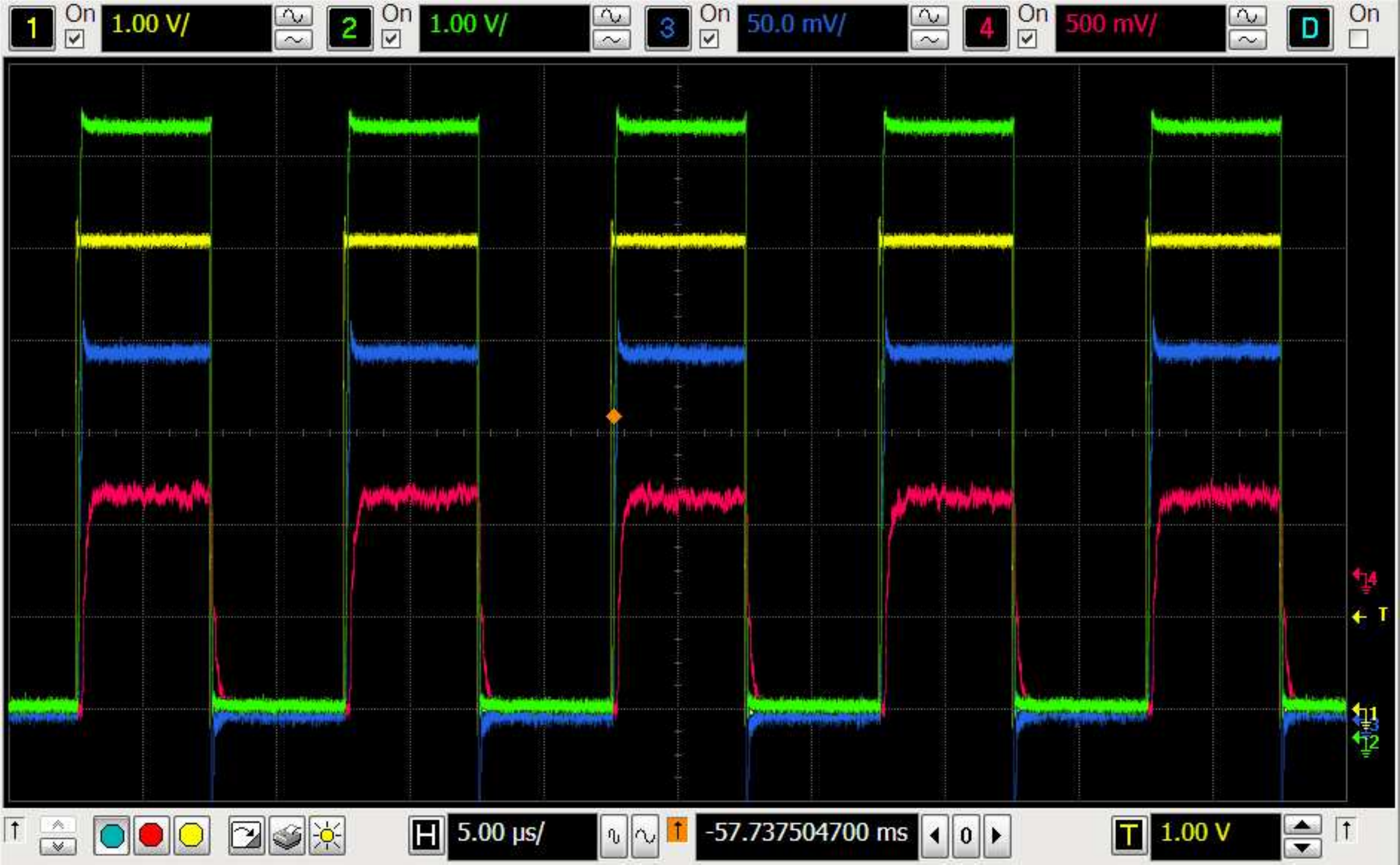}
\caption[Pulsed properties.]{\label{fig:PulsedProp} SET-255-01 pulsed at $100\,\textup{kHz}$, $50\%$ duty cycle at $20\,\textup{mA}$ drive current amplitude. The measurement was made at $20\,^{\circ}\textup{C}$ with a single pulse shown on the left and consecutive pulses on the right. Going from the top down, the first signal (green) is the voltage across the device, the second (yellow) is the initial pulse generator signal, the third (blue) is the current through the device (converted to a voltage at $10\,\textup{mV/mA}$) and the fourth (red) signal shows the amplified response of the PMT, where the rise/fall times are limited by the amplifier.}
\end{figure}

Pulsed capabilities were tested with the DUT mounted in the light sealed, temperature controlled enclosure. The UV output was routed to either the PMT or alternatively the calibrated power meter which allowed the average UV output to be measured. Each device was driven at a range of frequencies ($100\,\textup{Hz}$, $1\,\textup{kHz}$, $100\,\textup{kHz}$ and $1\,\textup{MHz}$ all at a $50\%$ duty cycle, $20\,\textup{mA}$ amplitude), duty cycles ($25\%$, $10\%$ and $5\%$ all at $100\,\textup{kHz}$, $20\,\textup{mA}$ amplitude) and current amplitudes ($10\,\textup{mA}$, $5\,\textup{mA}$ and $1\,\textup{mA}$ all at $100\,\textup{kHz}$, $25\%$ duty cycle). A full set of measurements were made at $-10\,^{\circ}\textup{C}$, $+20\,^{\circ}\textup{C}$ and $+40\,^{\circ}\textup{C}$ giving a total of 30 individual readings for each device. Example traces are shown in Figure \ref{fig:PulsedProp}.

The oscilloscope was used to measure the $10\%$ to $90\%$ rise time of the voltage across the DUT and it was found to scale linearly with the inverse of the drive current amplitude for all three device types. 
For the SET-255 devices the average rise time was $1.63\pm0.07\,\mu\textup{s}$ at $1\,\textup{mA}$ falling to $0.110\pm0.004\,\mu\textup{s}$ at $20\,\textup{mA}$ with the SET-240 devices showing similar behaviour with an average of $1.49\pm0.09\,\mu \textup{s}$ at $1\,\textup{mA}$ and $0.109\pm0.004\,\mu \textup{s}$ at $20\,\textup{mA}$. However, the CIS-250 devices were about four times slower over the current range studied with average rise times of $6.46\pm0.05\,\mu \textup{s}$ at $1\,\textup{mA}$ and $0.323\pm0.003\,\mu \textup{s}$ at $20\,\textup{mA}$. 

All three device types were able to be pulsed to greater than $100\,\textup{kHz}$ and due to their shorter rise times the SET devices up to $1\,\textup{MHz}$ with a $50\%$ duty cycle. It was also found that temperature had no measurable effect on any of the device modulation properties and at each temperature the average UV output power scaled linearly with pulse duration. As expected the average UV output at a particular drive current amplitude did vary with temperature as discussed Section \ref{TV}.

\subsection{Thermal Vacuum Performance}\label{TV}

\begin{figure}[h]
\begin{minipage}[t]{0.5\textwidth}
\centering
\includegraphics[width=1.0\textwidth]{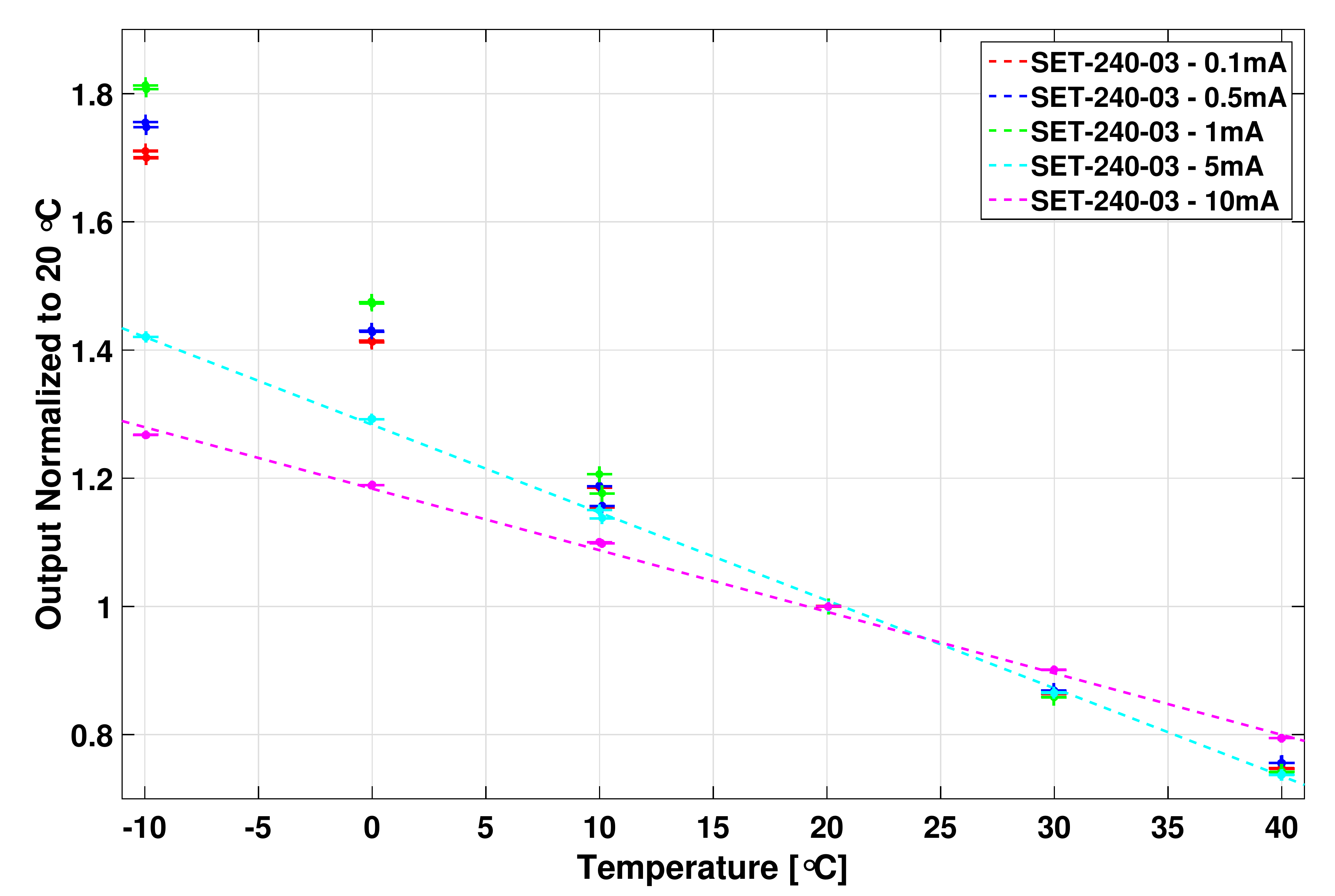}
\includegraphics[width=1.0\textwidth]{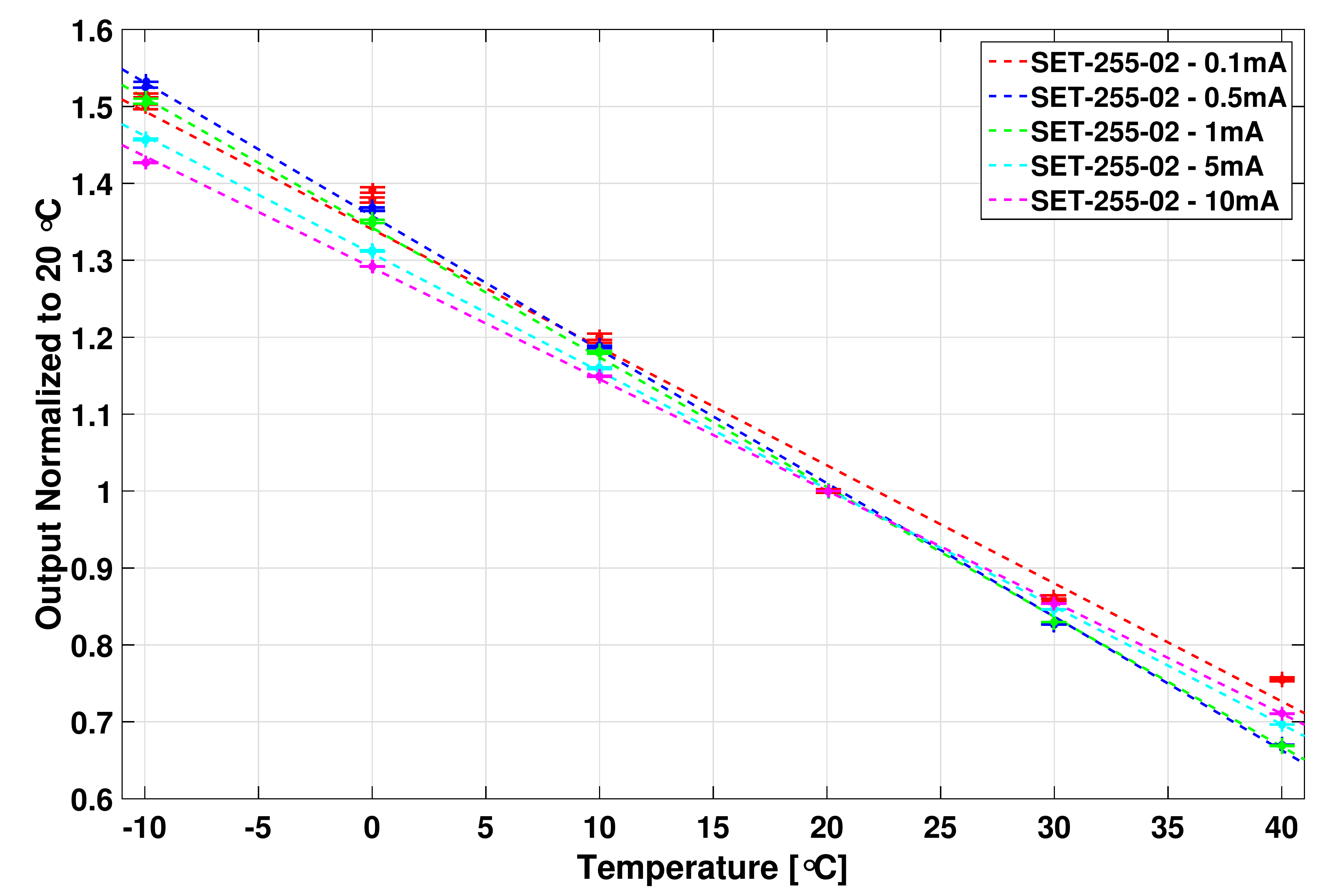}
\end{minipage}
\begin{minipage}[t]{0.5\textwidth}
\includegraphics[width=1.0\textwidth]{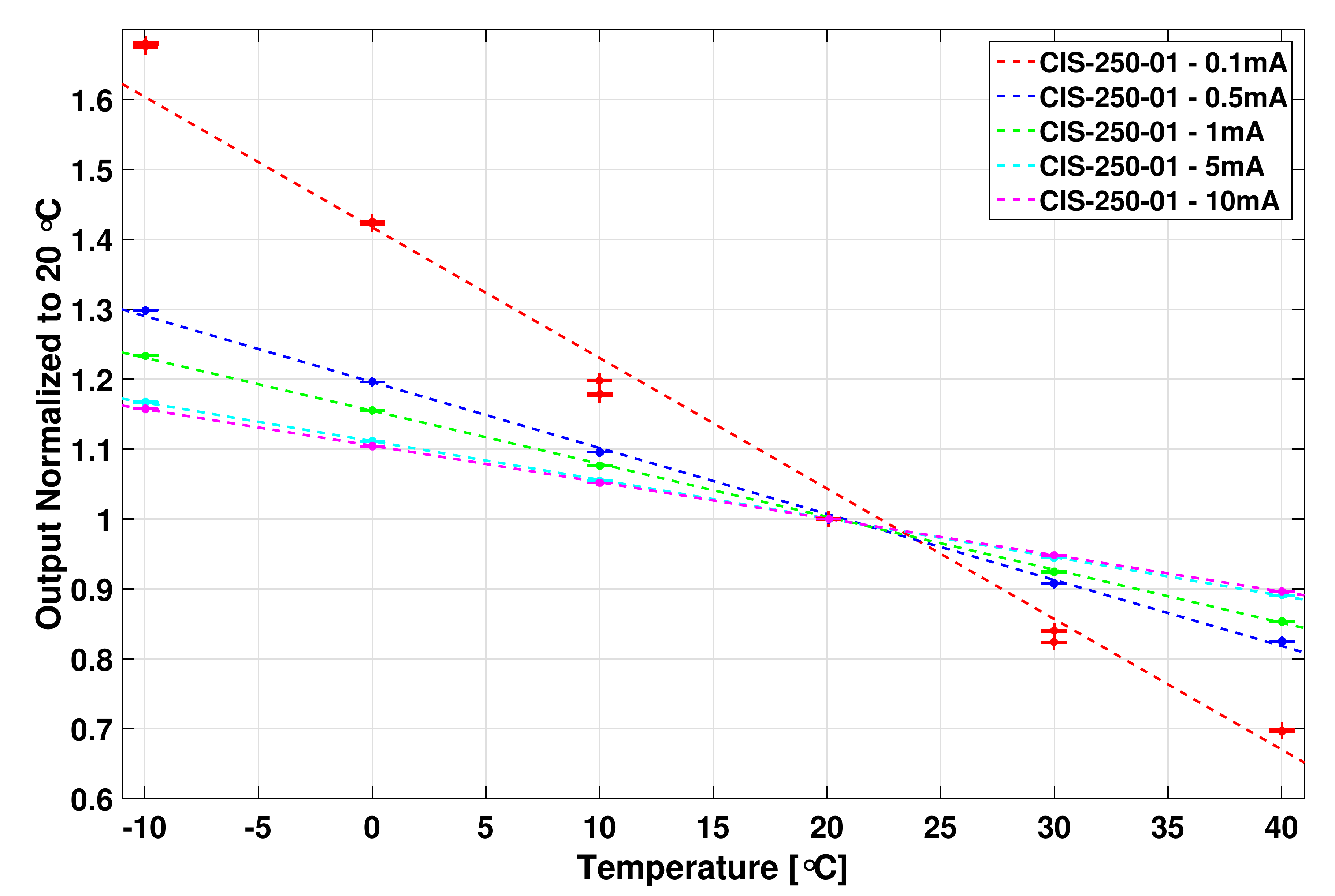}
\caption[UV output with varying temperature.]{\label{fig:OutputVsTemp} Variation in UV output between $-10\,^{\circ}\textup{C}$ and $+40\,^{\circ}\textup{C}$, at five different current settings. Note two separate readings were made at each current setting and each measurement has been normalized to the one made at $20\,^{\circ}\textup{C}$. Clockwise form bottom left: SET-255-02, SET-240-03 and CIS-250-01.}
\end{minipage}
\end{figure}

Thermal testing was carried out with a single device at a time mounted in a vacuum chamber at a pressure of $<10^{-5}\,\textup{mbar}$. The DUT was fixed in a copper mount with temperature control provided by a custom, two-stage Peltier system. The same type of Hamamatsu large area photodiode used previously was mounted opposite to the DUT to allow \textit{in situ} measurements of the total UV output. The first part of the test was designed to quantify the relationship between UV output and temperature, at a range of DC drive currents. The temperature was raised from $-10\,^{\circ}\textup{C}$ to $+40\,^{\circ}\textup{C}$ in steps of $10\,^{\circ}\textup{C}$, with this sequence repeated twice. Once the temperature had stabilised at each setting, an IV scan from $0\,\textup{mA}$ to $10\,\textup{mA}$ in steps of $0.1\,\textup{mA}$ was taken. In addition, the DUT was driven for 120 seconds in turn at $0.1\,\textup{mA}$, $0.5\,\textup{mA}$, $1\,\textup{mA}$, $5\,\textup{mA}$ and $10\,\textup{mA}$. The test was completely automated and lasted approximately 8 hours per device. The data were split by temperature and the average UV output was calculated from 40 seconds of data at each drive current setting. Data where the output was still settling were ignored. All nine devices were tested in this way with a selection of results shown in Figure \ref{fig:OutputVsTemp}.

As expected the UV output of all devices was found to have a significant temperature dependence which varied depending on the DC drive current. The behaviour was quantitatively similar for devices of a particular type but, as can be seen in Figure \ref{fig:OutputVsTemp}, differences were observed between types. The SET-255 devices were the most sensitive to temperature with a linear relationship seen which changed with current setting, though this begins to become non-linear at $0.1\,\textup{mA}$. Qualitatively similar results were seen with the CIS-250 devices but for the SET-240 devices the linear behaviour only held at $10\,\textup{mA}$ and $5\,\textup{mA}$. Considering just the $10\,\textup{mA}$ data where all devices demonstrated linear behaviour, the SET-255 devices showed a $1.5\,\%/^{\circ}\textup{C}$ change compared to $20\,^{\circ}\textup{C}$, the SET-240 devices $1.0\,\%/^{\circ}\textup{C}$ and the CIS-250 $0.5\,\%/^{\circ}\textup{C}$.

The second part of the thermal vacuum test was non-operational survival cycling carried out in the same system but with the DUT turned off. The temperature was cycled ten times between $-40\,^{\circ}\textup{C}$ and $+60\,^{\circ}\textup{C}$ with a $1\,\textup{hour}$ dwell at each extreme. The pressure was $<10^{-5}\,\textup{mbar}$ throughout and all nine devices survived without any sign of degradation.

\subsection{Repeated Reference Characteristics}

\begin{figure}[h]
\centering
\includegraphics[width=0.49\textwidth]{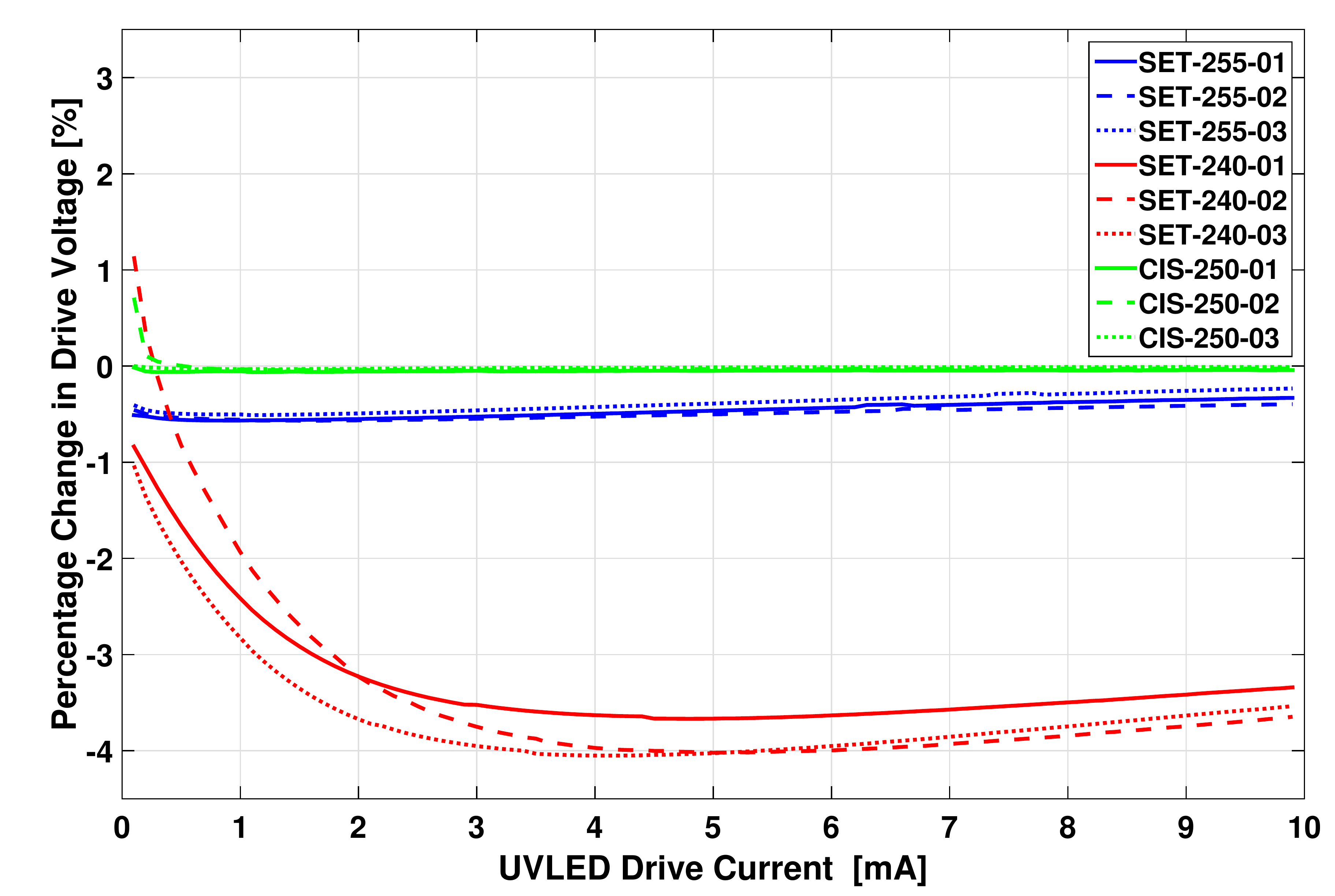}
\includegraphics[width=0.49\textwidth]{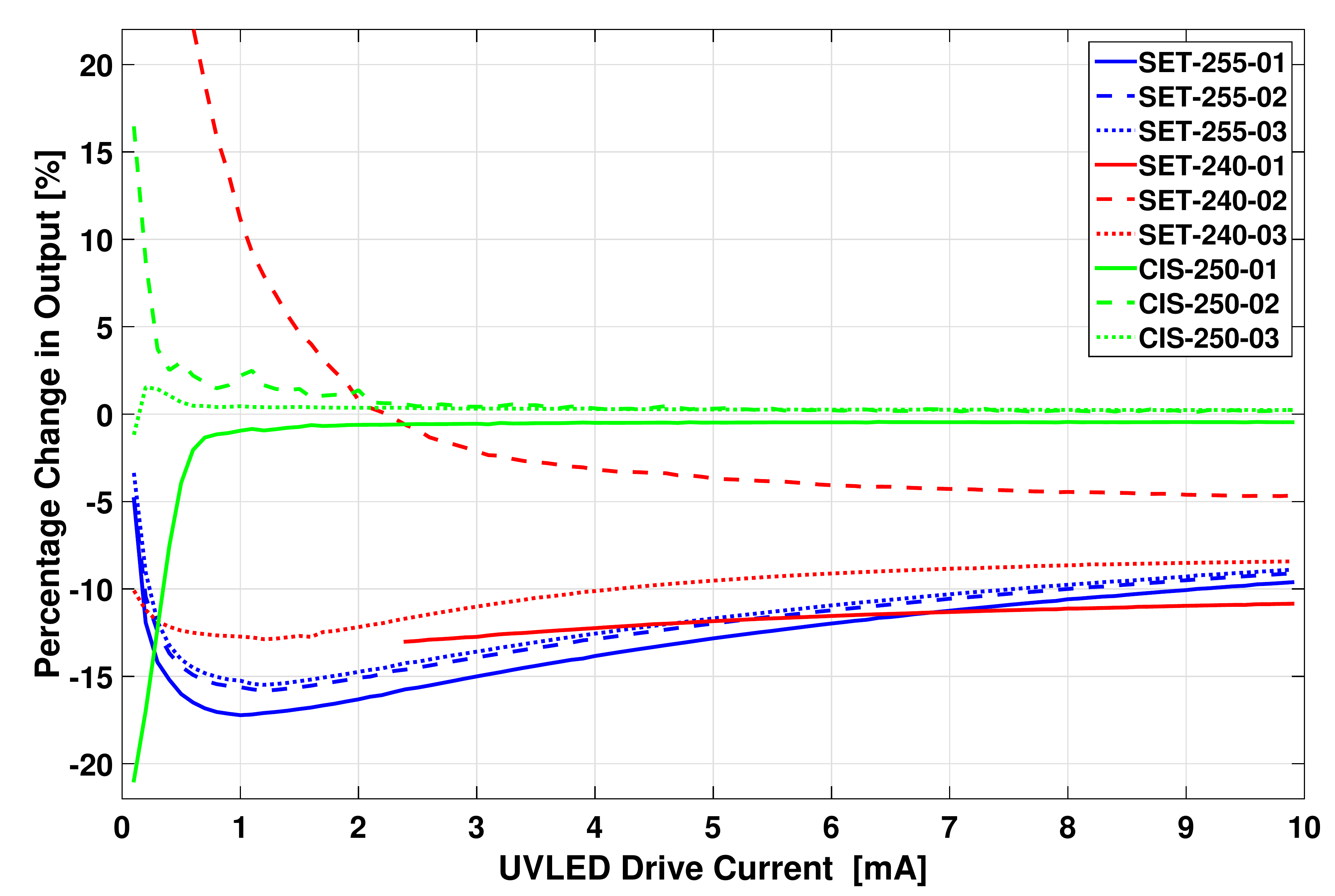}
\caption[Changes in device properties during laboratory tests.]{\label{fig:ChangePropLab} Left: The change in voltage required to drive a particular current. Right: The change in UV output at a particular drive current. Note the SET-240-01 data has been truncated below $2.2\,\textup{mA}$ to remove the data including the anomaly observed during the initial measurement. The SET-240-02 data rises steeply below $0.5\,\textup{mA}$ to approximately $+150\%$.}
\end{figure}

Following the laboratory-based testing, the reference measurements described previously were repeated. Within the measurement uncertainty, no device showed any change in spectral properties or pulsed behaviour. However, some changes were seen in both IV curves and UV output. The results are shown in Figure \ref{fig:ChangePropLab}.

Considering the IV properties first: CIS-250 devices showed no measurable changes; SET-255 devices experienced a reduction of $0.5\%$ to $0.25\%$ in drive voltage at all currents while the change for the SET-240 was larger and more complicated, though qualitatively similar for the three devices of this type. SET-240-02 stands out at lower currents as it transitions to an increase in required drive voltage.

Turning to the UV output, the CIS-250 devices again showed no significant change, at least at drive currents above $1\,\textup{mA}$ while the SET-255 devices experienced a reduction of $5\%$ to $15\%$ depending on the drive current. Two of the SET-240 devices showed similar behaviour but again SET-240-02 stands out. Below $2\,\textup{mA}$ its UV output appears to have increased significantly by up to $+150\%$ at drive currents below $0.5\,\textup{mA}$. Generally all devices showed more complex behaviour at lower drive currents, which was perhaps to be expected given the lower UV output. Recall that for all devices the total UV output power is around $100\,\mu\textup{W}$ at $10\,\textup{mA}$ and around $50\,\textup{nW}$ at $0.1\,\textup{mA}$. What is reassuring though is the repeatability of all the CIS-250 measurements as they provide confidence that the changes observed with the other devices are real.

\subsection{Radiation}

Following the laboratory-based tests the devices underwent radiation testing at the Cobalt-60 facility of the European Space Research and Technology Centre (ESTEC). A Co-60 source produced a diverging gamma-ray beam with photon energies of $1.17\,\textup{MeV}$ and $1.33\,\textup{MeV}$. The DUT dose rate was varied by adjusting the distance from the source. The total ionising dose and dose rate (water equivalent) was measured by a calibrated dosimeter provided by the facility. The test requirements called for a total dose of $30\,\textup{kRad}$ and this was delivered in three separate runs over the course of two days. The dose rates were adjusted to fit the test schedule and facility operating hours. For each of the device types one was irradiated while driven at $1\,\textup{mA}$ DC (01 devices), one was irradiated while off (02 devices) and one acted as a reference that was driven at $1\,\textup{mA}$ DC but was not irradiated (03 devices).

\begin{figure}[h]
\centering
\includegraphics[height=0.22\textheight]{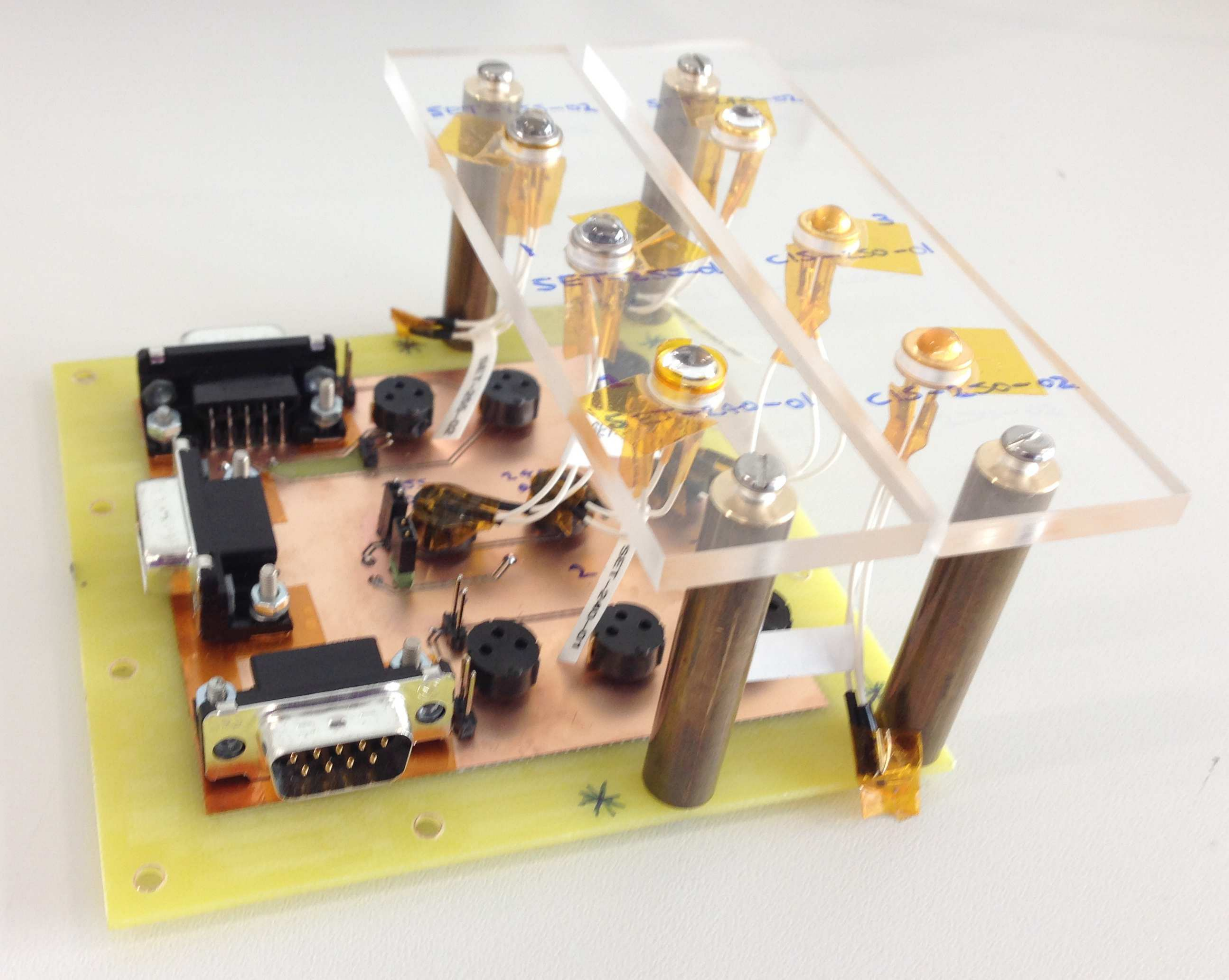}
\includegraphics[height=0.22\textheight]{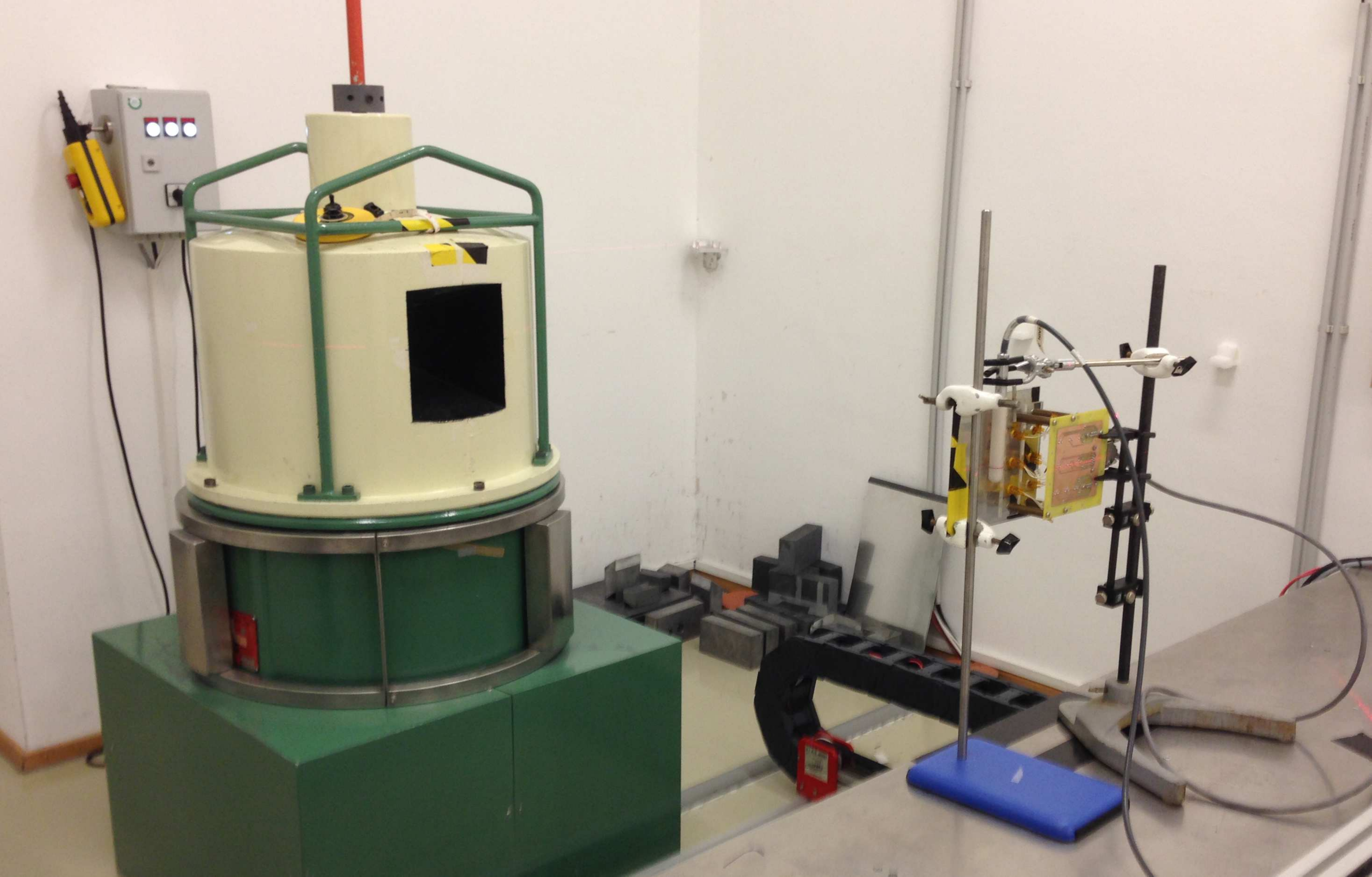}
\caption[Radiation testing.]{\label{fig:RadPhotos} Left: The six irradiated devices. Note that the leads of the unpowered devices are not connected to the board but have their lead tips covered with Kapton tape to avoid accidental electrical shorting. Right: The devices in place (to the right held in clamp stand) prior to irradiation.}
\end{figure}

Both the irradiated and reference devices were mounted within perspex holders with the flying leads of driven devices connected to a custom made printed circuit board (PCB), Figure \ref{fig:RadPhotos}. The devices that were driven were at a fixed $1\,\textup{mA}$ drive current, while the corresponding drive voltage was measured at $1/32\,\textup{Hz}$ sampling frequency by a data acquisition system provided by the facility. No active temperature control was used for the devices though the room temperature was controlled and measured to be between $23\,^{\circ}\textup{C}$ and $24\,^{\circ}\textup{C}$ throughout. There was no real-time monitoring of the UV output but IV scans as described previously were carried out in between irradiation runs.

The first run lasted 16 hours 45 minutes and delivered an absorbed dose of $13.26\,\textup{kRad}$, the second run lasted 3 hours 55 minutes and delivered an absorbed dose of $12.62\,\textup{kRad}$ and the third run lasted 16 hours 7 minutes and delivered an absorbed dose of $12.80\,\textup{kRad}$. The total absorbed dose in terms of water was $38.68\,\textup{kRad}$. 

\begin{figure}[h]
\centering
\includegraphics[height=0.22\textheight]{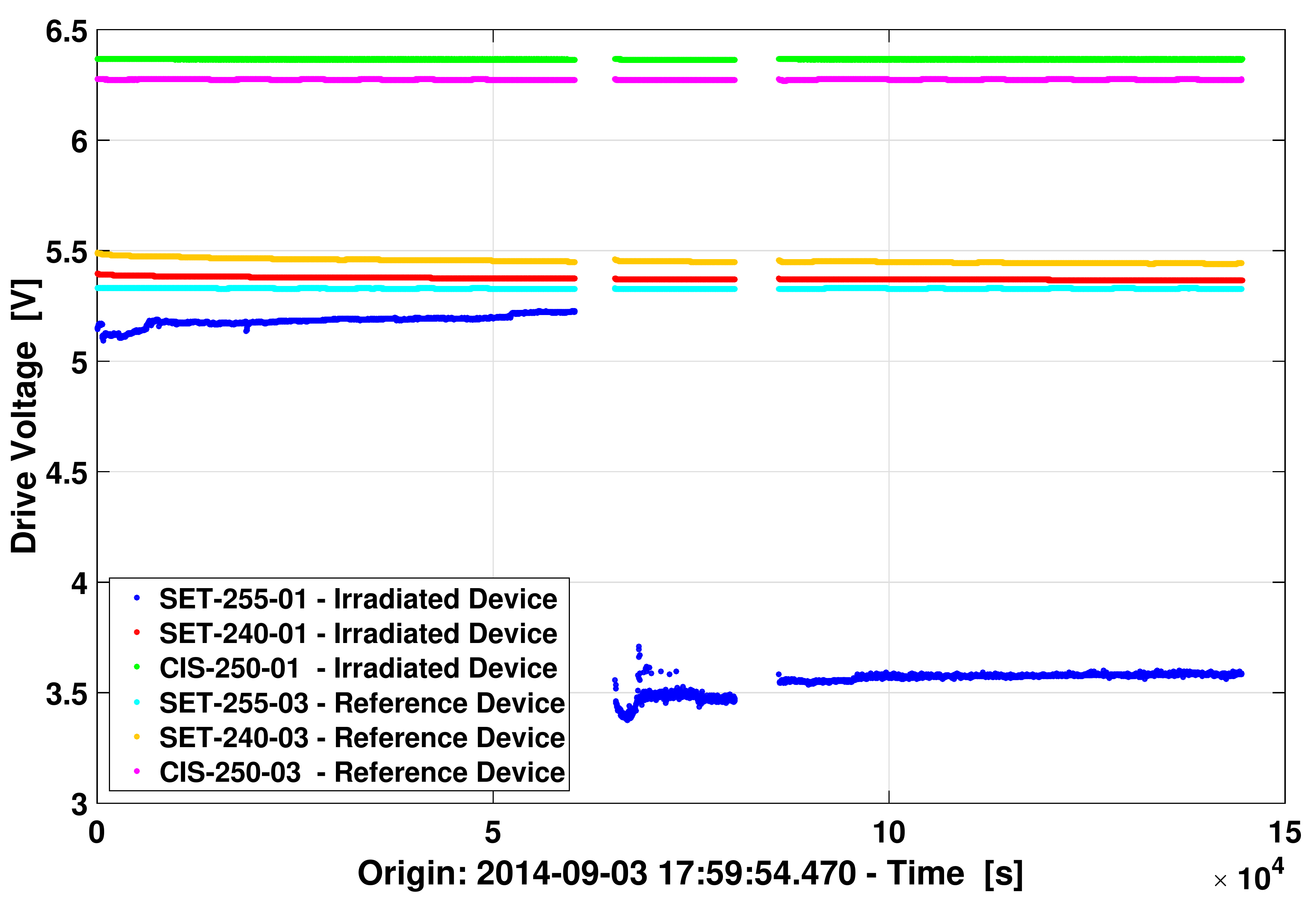}
\includegraphics[height=0.22\textheight]{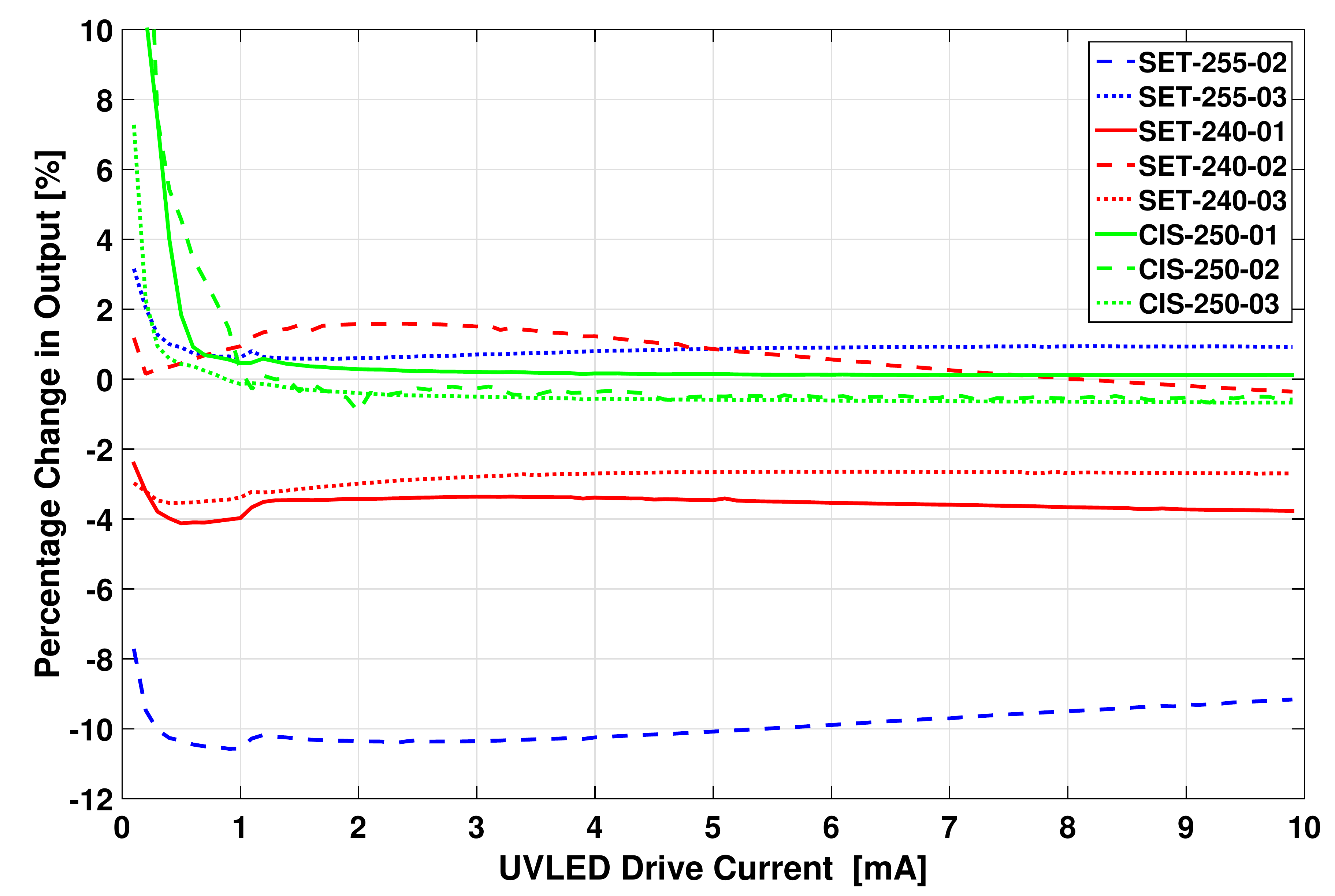}
\caption[Changes in device properties during radiation tests.]{\label{fig:ChangePropRad} Left: The monitoring voltages for the six devices that were driven at $1\,\textup{mA}$ during the test, three of which were also irradiated. SET-255-01 exhibited unstable behaviour during the first run and dropped by $1.75\,\textup{V}$ at the start of the second. Right: The change in UV output with drive current, as measured before and after the radiation test. Below $0.5\,\textup{mA}$, CIS-250-01 and CIS-250-02 rise to $12\%$ and $29\%$ respectively.}
\end{figure}

Soon after the test began it became clear from the drive voltage monitoring that SET-255-01 was behaving erratically, as can be seen in Figure \ref{fig:ChangePropRad}. This behaviour continued during the intermediate IV scans with the output of the device output changing by $\approx10\%$, although the device never completely failed. A subsequent inspection of data recorded upon arrival at the facility revealed that the drive voltage of the device was unstable prior to irradiation. This was in contrast to data recorded before transportation that showed no unusual behaviour. Thus, it would appear that the device could have been damaged either during transport or possibly during integration in the test rig and the subsequent behaviour was not caused by irradiation. More will be said about this in Section 5.7.

Following the test a full set of reference measurements were made at Imperial College, except for SET-255-01 which was too unstable. Within the uncertainty of the measurements, no changes were observed in either the spectral or modulation properties for any device and the IV scans showed no change in the drive voltage. Five out of the remaining eight devices also showed no significant  change in their UV output, though as can be seen in Figure \ref{fig:ChangePropRad}, three did. The device SET-255-02 which was irradiated but not driven showed a $\approx10\%$ reduction in UV output at all current settings. No such reduction was observed for the device SET-255-03 which was driven at $1\,\textup{mA}$ but not irradiated which suggests that the radiation exposure caused the degradation. The two SET-240 devices which were driven at $1\,\textup{mA}$ showed a $3\%$ to $4\%$ reduction in UV output at all current settings. However, device SET-240-02 which was irradiated but not driven showed no such reduction suggesting that ageing effects from the device being driven caused the degradation and not the radiation exposure. At high drive currents none of the CIS-250 devices showed any significant change though below $1\,\textup{mA}$ the output of all three increased.

\subsection{Vibration and Shock}

\begin{figure}[h]
\centering
\includegraphics[height=0.22\textheight]{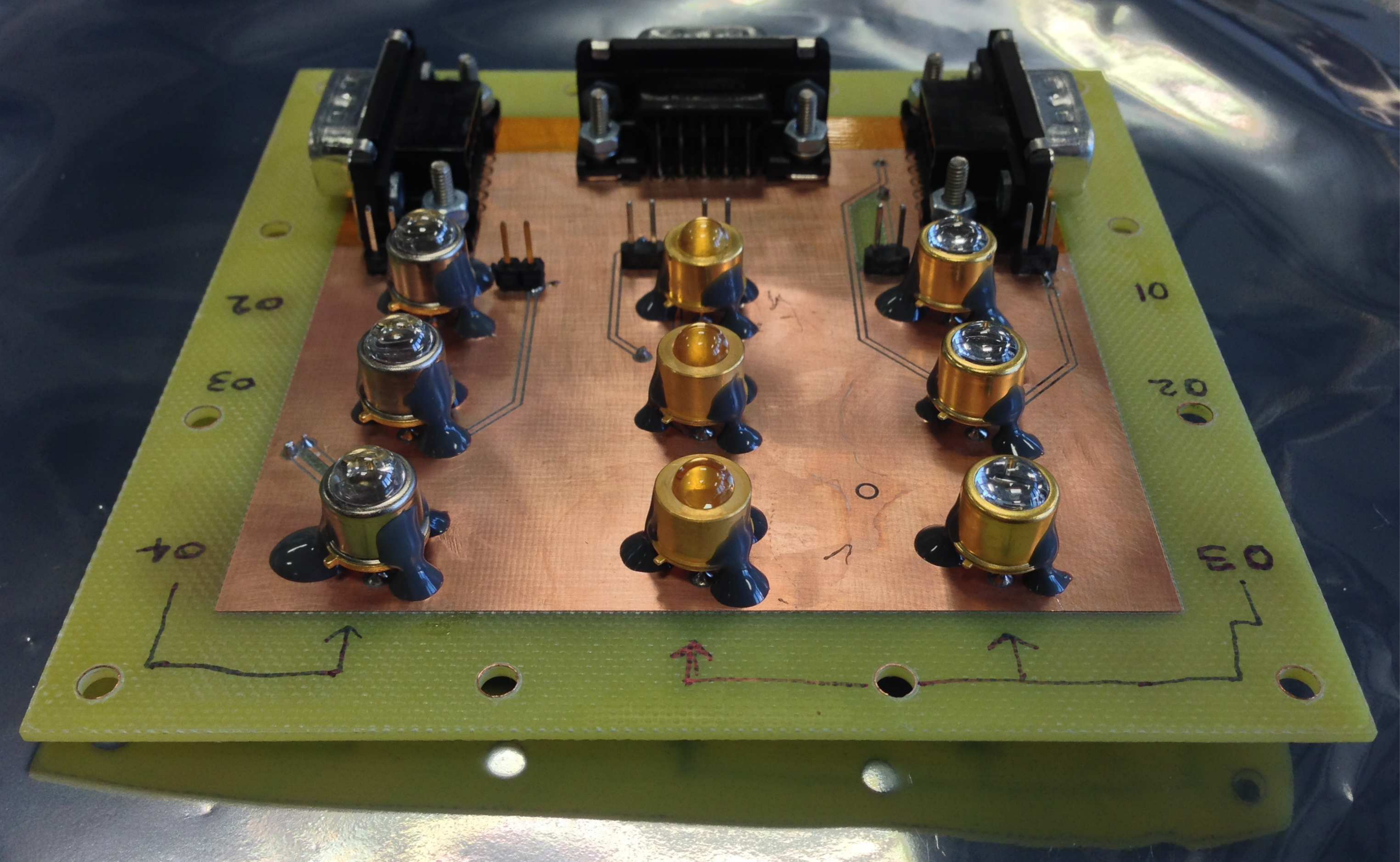}
\includegraphics[height=0.22\textheight]{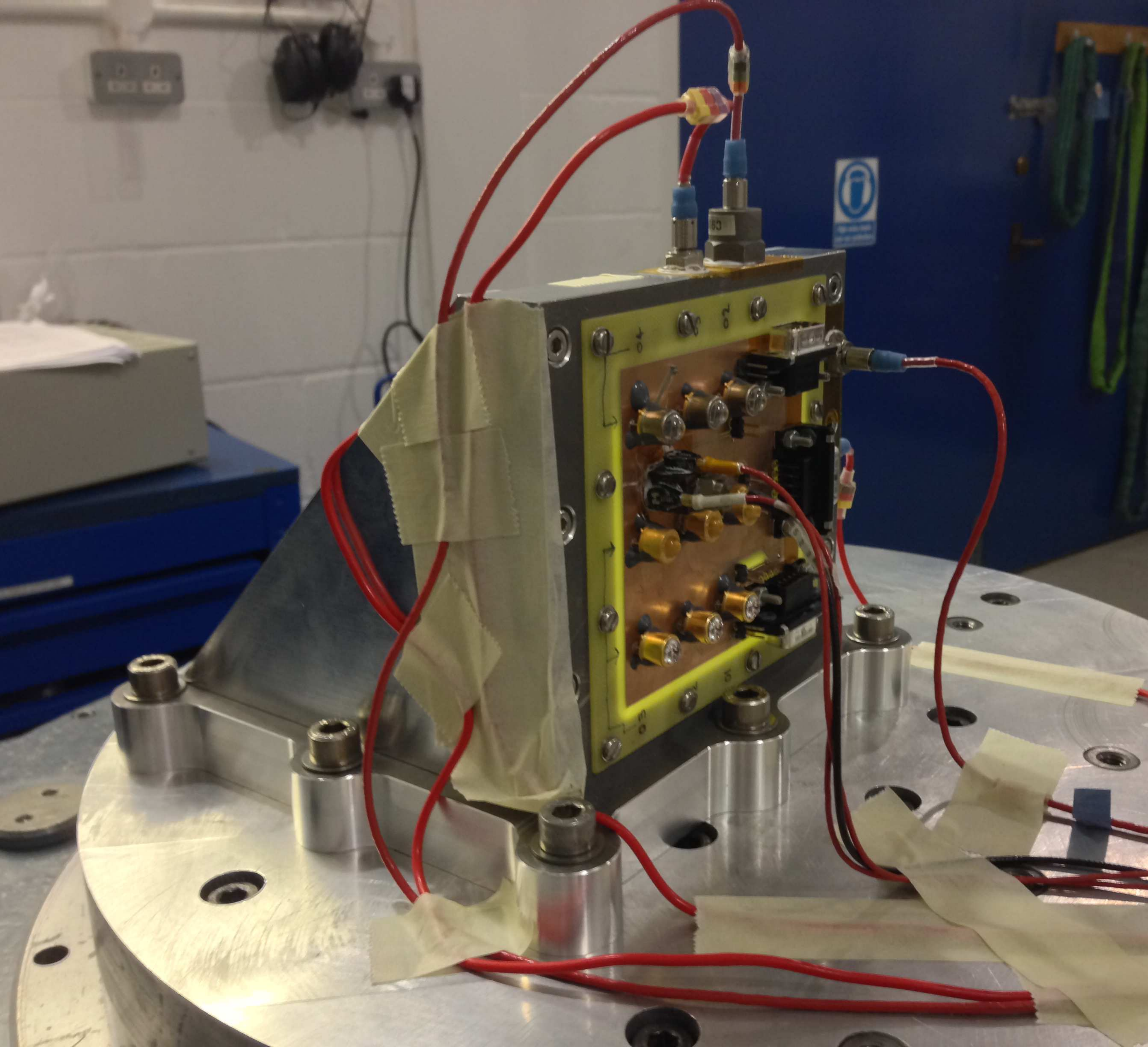}
\caption[Vibration testing.]{\label{fig:VibPhotos} Left: The nine devices soldered and staked to the test board, the SET-255 devices on the left, CIS-255 devices in the middle and the SET-240 devices on the right. Right: The test board mounted in the y-axis configuration with control and monitoring accelerometers visible. For the test in the x-axis the board was rotated $90^{\circ}$ and for the z-axis it was re-mounted such that the board was parallel to the base plate.}
\end{figure}

Vibration and shock testing was performed at the Airbus Defence and Space facility in Stevenage, UK. The device displaying erratic behaviour was replaced by a nominally identical one (SET-255-04) and all nine were soldered onto a custom made PCB according to ECSS including staking using Scotch-Weld 2216A/B (as can be seen in Figure \ref{fig:VibPhotos}) and strain relief \cite{ECSS2009}.

\begin{figure}[h]
\centering
\includegraphics[width=0.32\textwidth]{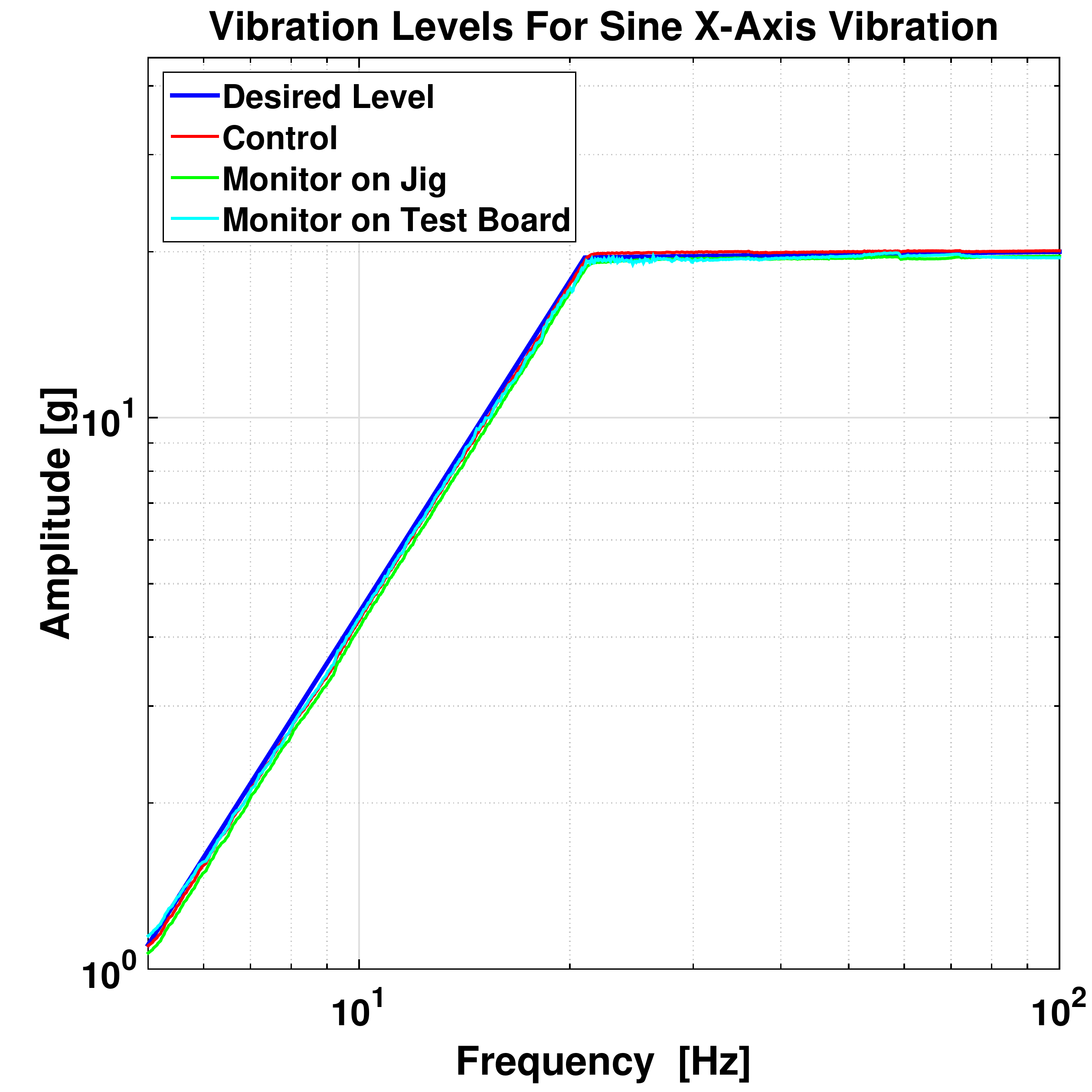}
\hfil
\includegraphics[width=0.32\textwidth]{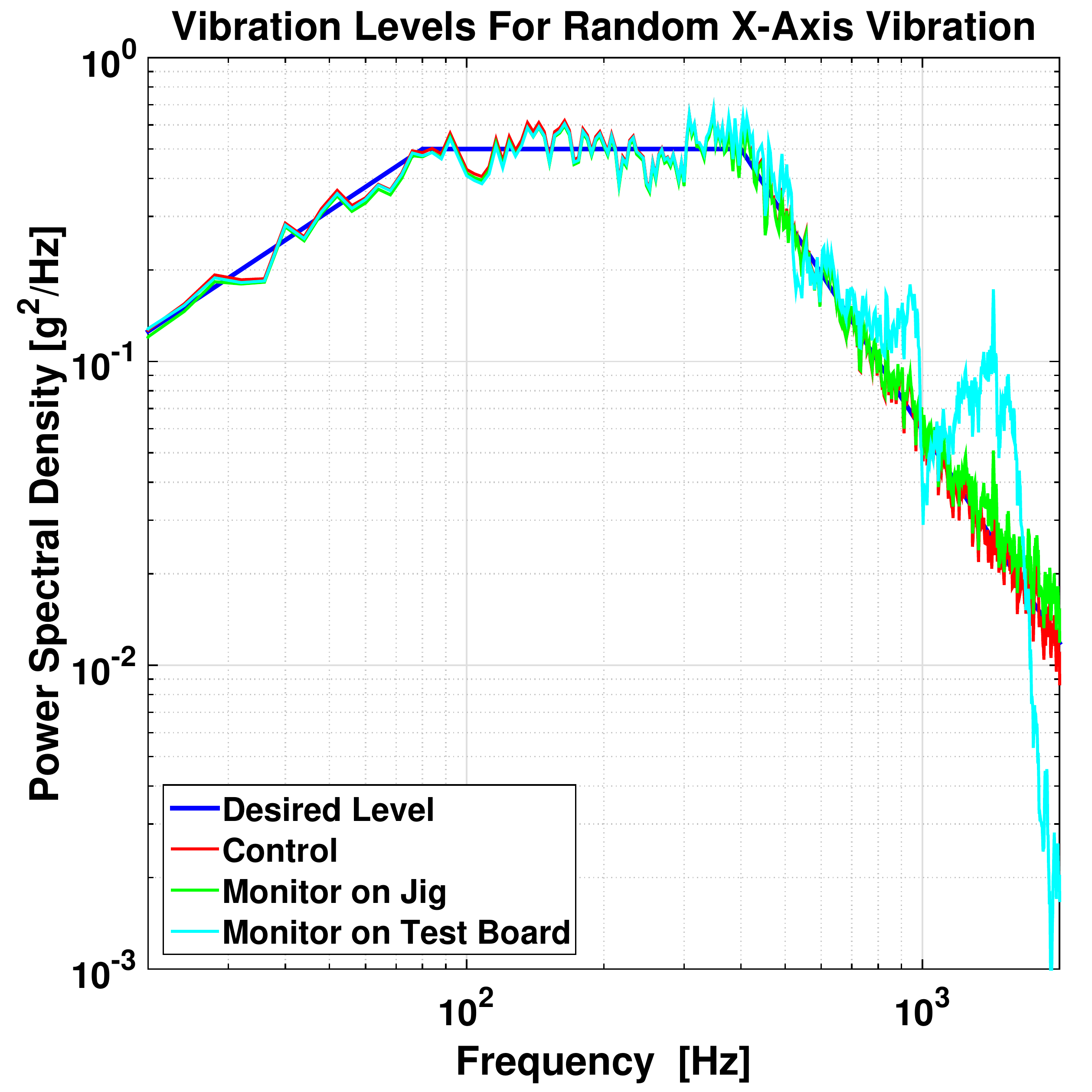}
\hfil
\includegraphics[width=0.32\textwidth]{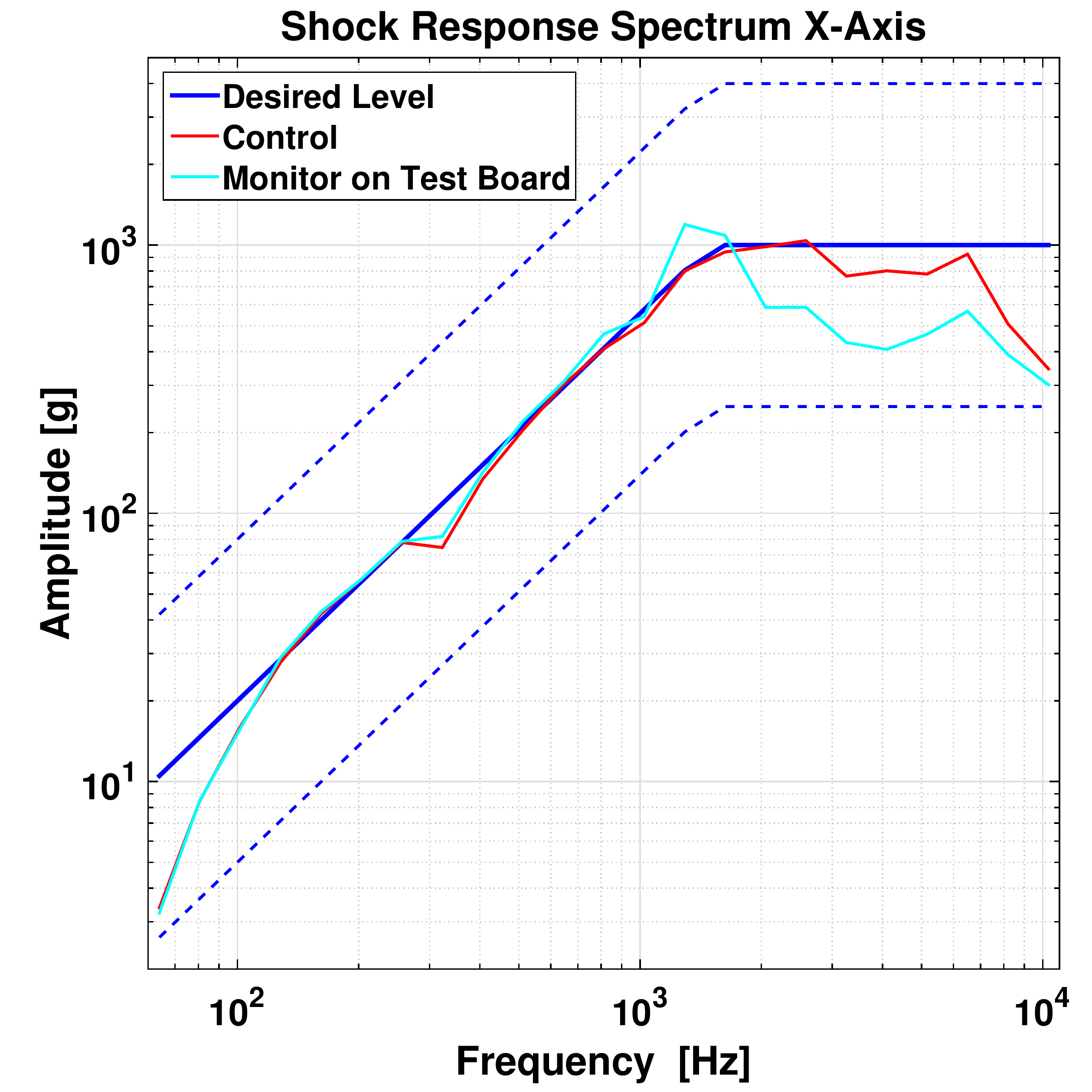}
\caption[Vibration levels.]{\label{fig:VibLevels} From left to right are the x-axis sine vibration at $11$\,mm 0--peak 20\,g, random $17.7$\,g rms and shock SRS test data. The tests were also carried out at the same levels in y and z.}
\end{figure}

Each axis was tested in turn by performing a quasi-static (In Plane: $\pm20\,\textup{g}$ at $35\,\textup{Hz}$ for 2 seconds. Out of Plane: $\pm30\,\textup{g}$ at $30\,\textup{Hz}$ for 2 seconds.) followed by a sine, random and shock test at the levels shown in Figure \ref{fig:VibLevels}, defined in \cite{Sumner2015}. In between each test a low level sine sweep was performed at $0.2\,\textup{g}$, from $5\,\textup{Hz}$ to $2\,\textup{kHz}$ at $2\,\textup{oct/min}$ to look for any changes.

Surprisingly, all three of the SET-255 devices suffered a complete failure during the first x-axis run each producing an open circuit during an inter-axis electrical turn on test. The three SET-240 and CIS-250 devices showed no change and successfully completed all tests. Upon return to Imperial College, all nine devices were carefully removed from the board and had their flying leads re-soldered. The reference tests were repeated and showed no significant change in either the spectral or modulation properties for any device. The IV scans showed no change in the voltage or UV output for any of the CIS-250 devices but some differences in the SET-240 properties. The drive voltage of all three devices had changed $\pm1\,\%$ over the $0\,\textup{mA}$ to $10\,\textup{mA}$ current range and the UV output of devices SET-240-02 and SET-240-03 had fallen by $\approx9\%$ and $\approx6\%$ respectively. It seems unlikely that the vibration test itself caused the changes but their are several other possibilities. For example, some degradation may have occurred due to a lack of heat sinking or temperature control while the devices were driven at $2\,\textup{mA}$ and $10\,\textup{mA}$ during the inter-axis electrical checks. Alternatively, the cleaning and re-soldering between reference tests could have led to measurable changes.

A visual inspection of the three failed SET-255 devices was performed with the aid of a microscope. No external damage was visible so an attempt was made to assess the inside of each device through the lens. With careful alignment this was indeed possible and revealed that the thin internal wire joining the chip to the device's legs had broken in both SET-255-02 and SET-255-04. The failure mode for SET-255-03 was less obvious but due to the difficult viewing conditions it remains possible that a more subtle break could have gone undetected. This points to insufficient internal stress relief and may also be contributing to the erratic behaviour of SET-255-01.

%%%%%%%%%%%%%%%%%%%%%%%%%%%%%%%%%%%%%%%%%%%%%%%%%%

\section{Conclusions}

The tests described here have shown that UV LEDs can offer superior performance in almost every way when compared to the mercury lamps employed in the LISA Pathfinder CMS. All three device types would be capable of producing more light than the mercury lamp system at wavelengths less than $254\,\textup{nm}$, with the SET-240 also emitting $\approx1\%$ at an energy greater than the work function of pure gold. Even in a simple DC drive scenario they would also offer a considerable improvement in dynamic range, as over the drive current range studied ($0.1\,\textup{mA}$ to $10\,\textup{mA}$) dynamic ranges of order $10^{4}$ were observed. 

Although the electrical properties of the devices were consistent with data sheet values, all three types had measured peak wavelengths at the upper limit of their quoted values. The spectra of the light emitted by the UV LEDs were shown to be very stable with DC drive current, pulsed duty cycle, irradiation and age. Temperature was shown to have a small effect on spectral FWHM and peak position but given the spacecraft will be very thermally stable by necessity, this is unlikely to be an issue.

It was also shown that all three device types can be pulsed to at least $100\,\textup{kHz}$ with a $50\%$ duty cycle, allowing them to be synchronised with the injection bias that will be present in the inertial sensor. As outlined, the ability to synchronise with the injection bias offers several additional advantages including mitigating the risk of asymmetric surface properties, increasing dynamic range and improving the efficiency of discharging. Pulsed performance was found to be stable with temperature but the rise times of all devices varied with the inverse of the drive current amplitude. With respect to rise times, both SET devices were similar but the CIS devices were found to be around four times slower over the current range studied.

All nine devices survived thermal vacuum cycling and were shown to operate over a temperature range of at least $-10\,^{\circ}\textup{C}$ to $+40\,^{\circ}\textup{C}$. As expected, the UV output of all device types was significantly temperature dependant and this relationship was also found to vary with DC drive current. Additionally, all devices were found to be radiation hard, though a possible $\approx10\%$ reduction in UV output was observed in SET-255-02. During this test SET-255-01 began to behave erratically but there is evidence to suggest the behaviour was not caused by the irradiation, though it may have exasperated the problem.

Both the SET-240 and CIS-250 devices survived vibration and shock testing in all three axes. However, unexpectedly all three of the SET-255 devices suffered a complete failure after the first x-axis test. Further investigation revealed that at least two of the three devices failed due to the breakage of a thin wire bond within the TO-39 package. As all three device types appear to have similar internal mechanical structure it is not clear why only the SET-255 devices would fail in this way. It is possible that the devices were just part of a bad batch or maybe the issue relates to the integrated ball lens, which was the largest of the three types studied. On a positive note the failure was not due to the underlying UV chip technology so a change in internal design could mitigate any risk.

The UV output of both types of SET device gradually degraded over the course of the tests, despite the fact usage was fairly minimal and always within data sheet recommendations. As measured at the end of the radiation tests the UV output of the SET-255 devices had fallen by an average of $\approx15\%$ at drive currents above $1\,\textup{mA}$. The SET-240 devices were similar with an average fall of $\approx13\%$ at drive currents above $1\,\textup{mA}$. This is in contrast to the CIS-250 devices where the measured change was $<1\%$ at drive currents above $1\,\textup{mA}$. Interestingly, all nine devices showed significant changes in UV output below approximately $1\,\textup{mA}$ where increases and decreases of tens of percent were observed for devices of each type.

Each device accumulated an average total usage of $30\,\textup{mA hours}$ during testing, with the devices that were driven during the radiation test (devices 01 and 03) gaining an additional $37\,\textup{mA hours}$. This can be considered a low-level of usage as according to the manufacturers, the CIS-250 devices receive a burn-in of $4800\,\textup{mA hours}$ ($48\,\textup{hours}$ at $100\,\textup{mA}$) prior to delivery. This burn-in may go some way to explaining the apparent output stability of the CIS-250 devices, at least at higher DC currents. Nevertheless the results do add support to the manufacturers claim that the technology used in their devices offer superior lifetimes.

With a few caveats, it has been demonstrated that the two underlying technologies behind the three devices studied are suitable for use in space. When compared directly there are several pros and cons for all three device types tested, with ultimately there being a trade off between shorter wavelengths, faster rise times and reliability. The study has also highlighted that thermal management and monitoring will be an important design consideration in the final charge management system and that a cautious approach should be employed with regards to vibration. Initial results from ongoing device lifetime testing have also been positive and will be reported separately \cite{Hollington2015}.

%%%%%%%%%%%%%%%%%%%%%%%%%%%%%%%%%%%%%%%%%%%%%%%%%%

\section*{Acknowledgements}

The authors would like to thank the whole LTPDA software development team \cite{LTPDA2015} as well as Michele Muschitiello at ESTEC and Jaime Fensome at Airbus Stevenage for their assistance during the radiation and vibration tests respectively. Our thanks also go to Shahid Hanif at Imperial College for designing and building the pulsed drive electronics. We also acknowledge the financial support of the European Space Agency under contract C4000103768 for this work.

%%%%%%%%%%%%%%%%%%%%%%%%%%%%%%%%%%%%%%%%%%%%%%%%%%

\section*{References}

\bibliographystyle{unsrt}
\bibliography{bibliography}

\end{document}